\documentclass[sigconf]{acmart}

\usepackage[ruled,vlined]{algorithm2e}
\usepackage{amsmath}
\usepackage{graphicx}
\usepackage{subfigure}
\usepackage{caption}

\usepackage{subcaption}

\usepackage{multirow}
\usepackage{makecell}
\usepackage{threeparttable}
\usepackage{tablefootnote}

\AtBeginDocument{%
  }

\setcopyright{acmlicensed}
\copyrightyear{2025}
\acmYear{2025}

\usepackage{xcolor}
\usepackage{enumitem}
\usepackage{ulem}

\begin{document}


\title{Speech Token Prediction via Compressed-to-fine Language Modeling}

\author{Wenrui Liu, Qian Chen, Wen Wang, Yafeng Chen, Jin Xu, Zhifang Guo, Guanrou Yang, Weiqin Li, Xiaoda Yang, Tao Jin, Minghui Fang, Jialong Zuo, Bai Jionghao, Zemin Liu\footnotemark[2] \\ Zhejiang University}
\email{liuwenrui@zju.edu.cn}



\begin{abstract}
Neural audio codecs, used as speech tokenizers, have demonstrated remarkable potential in the field of speech generation. However, to ensure high-fidelity audio reconstruction, neural audio codecs typically encode audio into long sequences of speech tokens, posing a significant challenge for downstream language models in long-context modeling. We observe that speech token sequences exhibit short-range dependency: due to the monotonic alignment between text and speech in text-to-speech (TTS) tasks, the prediction of the current token primarily relies on its local context, while long-range tokens contribute less to the current token prediction and often contain redundant information. Inspired by this observation, we propose a \textbf{compressed-to-fine language modeling} approach to address the challenge of long sequence speech tokens within neural codec language models: (1) \textbf{Fine-grained Initial and Short-range Information}: Our approach retains the prompt and local tokens during prediction to ensure text alignment and the integrity of paralinguistic information; (2) \textbf{Compressed Long-range Context}: Our approach compresses long-range token spans into compact representations to reduce redundant information while preserving essential semantics. Extensive experiments on various neural audio codecs and downstream language models validate the effectiveness and generalizability of the proposed approach, highlighting the importance of token compression in improving speech generation within neural codec language models. The demo of audio samples will be available at \url{https://anonymous.4open.science/r/SpeechTokenPredictionViaCompressedToFinedLM}.

\end{abstract}




\maketitle

\section{Introduction}
Recent breakthroughs in neural codec language models have significantly advanced the field of speech generation, enabling the generation of more natural and high-quality speech in zero-shot TTS task~\citep{valle, soundstorm, maskgct, halle} and voice-interactive scenarios~\cite{speechgpt, moshi, vita, minmo, baichuanomni}. Neural codec language models translate input text into a sequence of speech tokens, which are then decoded into the target speech waveform by neural audio codecs~\citep{encodec, soundstream, hubert, bigcodec, speechtokenizer, wavtokenizer}. Existing neural codec language models are built upon diverse architectures, including autoregressive decoders~\citep{xtts, wavtokenizer}, non-autoregressive decoders~\cite{maskgct}, and hybrid transformers~\citep{valle, ellav, valler, valle2, halle}. 


To ensure high-fidelity audio reconstruction, neural audio codecs often discretize speech waveform into long sequences of speech tokens. Although current research trends favor using neural audio codecs with lower frame rates~\cite{ticodec, bigcodec, wavtokenizer} and fewer codebooks~\cite{hificodec, halle}, encoding a speech segment of 7-8 seconds still requires hundreds or even thousands of speech tokens. The excessive length of the speech token sequences presents challenges to long-context modeling, leading to the following critical issues:



\textbf{Sparsity undermines local alignment.} To convey the same semantic information, speech token sequences are typically much longer than text token sequences~\cite{wang2025vocalnetspeechllmmultitoken}, resulting in both \textit{semantic information sparsity} and \textit{attention sparsity}. ~\cite{kavaki2025audio} suggests that audio and speech exhibit information sparsity and that long-term dependency, which can be captured by transformer models, may not be critical for audio applications: attention between distant time frames may be deemed unnecessary~\cite{kim2020t}. Furthermore, attention sparsity is widely observed in transformer-based language models, where more tokens in long sequences dilute the attention distribution~\cite{shi2023large, tikochinski2025incremental, lu2025moba}. These two forms of sparsity converge in neural codec language models: (1) semantic information sparsity arises from the imbalance between text and speech token lengths; (2) attention sparsity stems from lengthy speech token sequences. The sparsity distracts attention weights between speech tokens and corresponding text tokens, and hinders the language model to capture local fine-grained alignment between text and speech.

\textbf{Redundancy introduces semantic noise}. Linguistic research found that acoustic redundancy naturally exists in speech and is closely related to duration and prosody~\cite{aylett2006language, jurafsky2008probabilistic, malisz2018dimensions}. ~\cite{dieleman2021variable, ticodec, sicherman2023analysing} demonstrate that this redundancy persists in discrete representations derived from speech signals, manifesting as repetitive timbres~\cite{ticodec}, silences~\cite{dieleman2021variable}, inconsistent units representing the same phoneme~\cite{sicherman2023analysing} and so on. Such redundant information is irrelevant to the current text, contributes less to the prediction of the current speech token and may even pose semantic noise, hence reducing the robustness of the neural codec language model~\cite{sicherman2023analysing}.


The mapping from text to speech inherently relies on local alignment between text and speech tokens, as exemplified by the monotonic alignment between phonemes and acoustic features~\cite{glowtts, radtts}. However, as explained above, sparsity and redundancy in long sequences of speech tokens disrupt the alignment process, leading to reduced prediction accuracy of speech tokens and finally degraded TTS performance. To address these challenges in neural codec language models, two primary approaches have been explored:

\textbf{(1) Ultra high-compression-rate speech tokenizers}~\cite{cho2024sylber, syllablelm} encode speech into syllable-like units, dynamically reducing token rates from 50Hz~\cite{hubert} to 4Hz-5Hz to preserve the main semantic information. However, high compression at the level of neural audio codecs risks losing acoustic details such as prosody and timbre characteristics, which are essential for high-quality speech reconstruction and generation.



\textbf{(2) Language modeling for robust speech token generation}, including group codec modeling~\cite{valle2} and forced alignment between text-speech tokens~\cite{valler, ellav}. UniAudio~\cite{uniaudio} and Mars6~\cite{mars6} employ the local-global-local modeling approach, which compresses and recovers speech tokens at the level of language models. However, these methods depend on particular architectures to mitigate sparsity, but redundancy remains inadequately addressed.

To address both sparsity and redundancy challenges and overcome limitations of prior approaches, in this paper, we propose an innovative and general solution for long-context modeling in neural codec language models. Our main contributions are as follows:


\begin{itemize}[leftmargin=*,noitemsep]

\item We analyze \textbf{the phenomenon of short-range dependency} in neural codec language models, where prompt and local tokens provide necessary information for language models to predict speech tokens. We find that relying solely on prompt and local tokens results in 2\% absolute reduction in WER and 2\% absolute gain in speaker similarity on TTS task, compared to the method of leveraging all previous speech tokens. 

\item Inspired by this finding, we introduce the \textbf{compressed-to-fine language modeling approach}, which compresses speech tokens at the level of language models, and overcomes the limitations of token compression solely at the level of neural audio codecs. Our approach serves as an information bottleneck by compressing long-range spans into compact representations and retains local tokens, effectively filtering out long-range redundant information and also alleviating the sparsity issue.


\item We validate the effectiveness and generalizability of the proposed method across representative and competitive neural audio codecs (e.g., EnCodec~\cite{encodec}, WavTokenizer~\cite{wavtokenizer}) and neural codec language models (e.g., VALL-E~\cite{valle}, decoder-only transformer). Experimental results demonstrate that our method improves VALL-E by 2.84\% absolute reduction in WER and 2.89\% absolute improvement in speaker similarity. When using WavTokenizer as speech tokenizer and decoder-only language model for TTS, applying the proposed method also achieves 2.5\% absolute and 42.8\% relative reduction in WER.
\end{itemize}

\section{Related Work}

\noindent \textbf{Neural Audio Codecs.} Discrete speech tokens are typically extracted from continuous audio signals, by speech models or neural audio codecs. HuBERT~\cite{hubert} and $S^3$ tokenizer~\cite{funaudiollm, cosyvoice} extract semantic tokens via K-means or Vector Quantization (VQ). EnCodec~\cite{encodec}, HiFiCodec~\cite{hificodec}, SpeechTokenizer~\cite{speechtokenizer}, and Xcodec~\cite{xcodec} employ Residual Vector Quantization (RVQ) for hierarchical multi-codebook discretization, while WavTokenizer~\cite{wavtokenizer} and UniCodec~\cite{unicodec} use a single codebook to encode all acoustic features. These codecs share a critical limitation: encoding 7–8 seconds of audio into hundreds of tokens introduces not only computational inefficiencies~\cite{valle2, valler}, but also long-context modeling challenges~\cite{valle2, lu2025moba} such as sparsity~\cite{shi2023large, tikochinski2025incremental, lu2025moba} and redundancy~\cite{dieleman2021variable, ticodec}. ~\cite{cho2024sylber, syllablelm} compress sequences of speech representations from 50Hz~\cite{hubert} into 4Hz–5Hz syllable-like units but lose acoustic details needed for high-quality reconstruction and synthesis. Prior works expose a core trade-off: \textit{increasing sequence length but introducing the issue of long-context modeling, or shortening sequence length at the cost of losing acoustic details.}

To address this dilemma, \textbf{our solution shifts compression from the level of neural audio codecs to the level of neural codec language models}, retaining the original codec's high-fidelity reconstruction through raw tokens while processing compressed 5Hz semantics-like units, a granularity shown to provide sufficient semantic information~\cite{guo2025recent, cho2024sylber, syllablelm}.

\noindent \textbf{Neural Codec Language Models.} Neural codec language models predict speech tokens, which are then decoded into continuous waveforms by neural audio codecs. VALL-E~\cite{valle} employs EnCodec~\cite{encodec} to encode speech and predicts speech tokens through a coarse-to-fine approach. VALL-E R~\cite{valler} introduces durations as additional labels, enforcing monotonic alignment between speech tokens and phonemes. VALL-E 2~\cite{valle2} employs group codec modeling to predict multiple tokens (typically 2 tokens) in a single autoregressive (AR) iteration. USLM~\cite{speechtokenizer} adopts SpeechTokenizer, which decouples semantics and acoustics. 
UniAudio~\cite{uniaudio} adopts MegaByte~\cite{megabyte} to enhance its ability to process long sequences of speech tokens. VoiceCraft~\cite{voicecraft} optimizes the processing of multi-codebook information through token rearrangement. 

While these methods enhance robustness of speech token prediction, they rely on \textit{specialized architectures} for neural audio codecs~\cite{speechtokenizer, halle} or language models~\cite{uniaudio, mars6, valler, halle, maskgct}, limiting their generalizability across diverse codecs and neural codec language models. Thus, \textbf{we propose compressed-to-fine language modeling, aiming at providing a universal approach for speech token prediction across various neural audio codecs and neural codec language models}.


\section{Methodology}

\begin{figure*}[t!]
    \centering
    \includegraphics[scale=0.36]{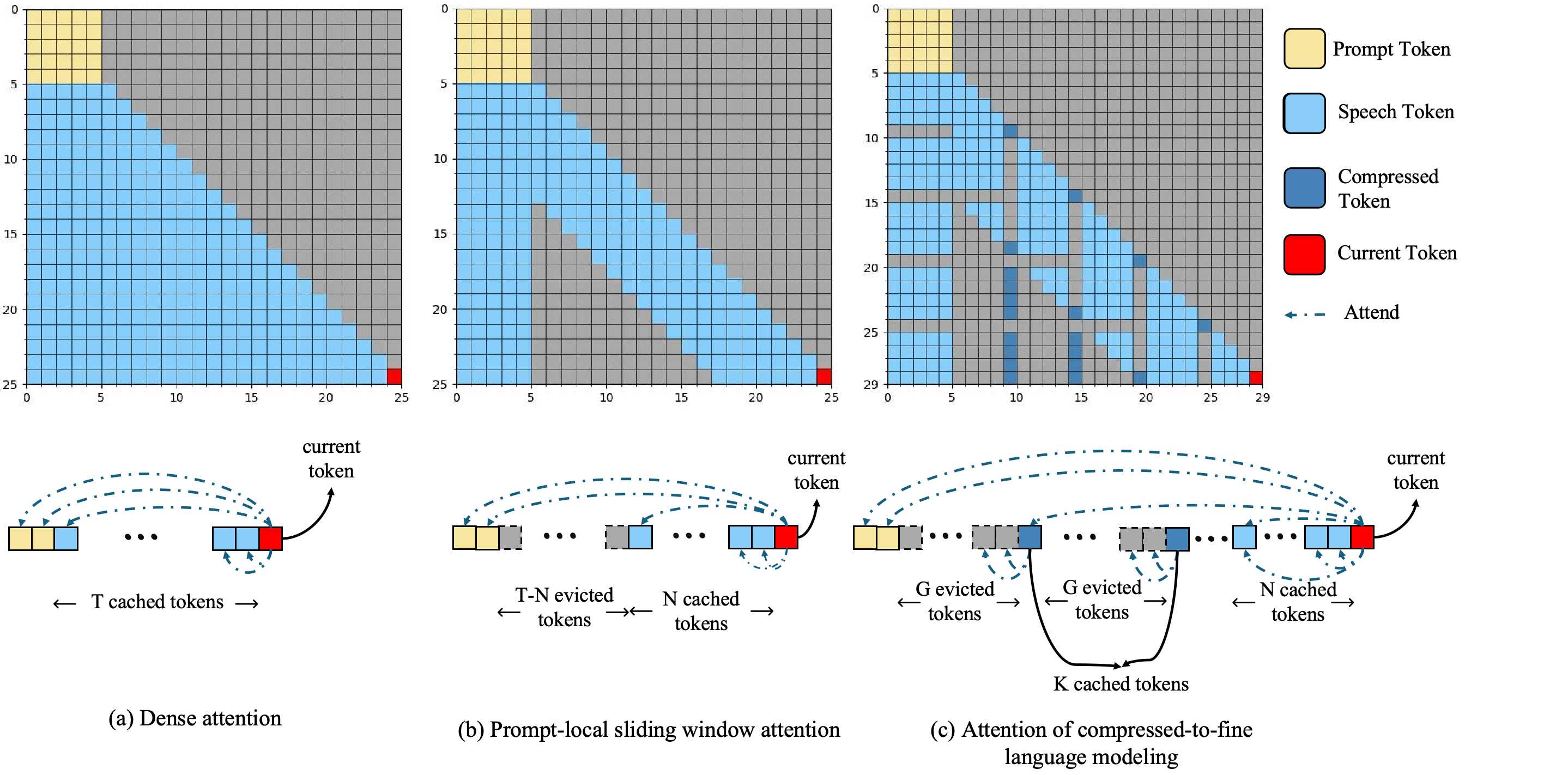}
    \caption{Illustration of the proposed Compressed-to-fine Language Modeling versus existing methods in autoregressive language models. (a) In \textit{dense attention}, the current speech token can attend to all previous prompt tokens and speech tokens. (b) In \textit{prompt-local sliding window attention}, the current speech token can only attend to prompt and local tokens within the sliding window. (c) In \textit{our compressed-to-fine language modeling}, the current speech token can attend to prompt tokens, compressed tokens of long-range spans, and local tokens. Illustration of the proposed method for non-autoregressive language models is provided in supplementary material.}
    \label{fig:attention}
\end{figure*}

\subsection{Preliminaries}
Neural codec language models typically follow a three-stage paradigm to generate speech:
(1) The neural audio codec encodes the speech prompt $S$ into a sequence of discrete speech tokens $C^{\text{\text{ref}}}$.  
(2) Based on the text prompt $T$ and speech prompt $C^{\text{\text{ref}}}$, the downstream language model predicts the sequence of speech tokens $C$.  
(3) The neural audio codec decodes $C$ into a continuous audio signal, which is the synthesized speech.  

Taking VALL-E~\cite{valle} as an example, neural codec language models generally consist of two components:
(1) An autoregressive decoder $\text{LM}_{\text{AR}}$, which predicts the first-layer codewords $C^{1}$.  
(2) A non-autoregressive decoder $\text{LM}_{\text{NAR}}$, which iteratively predicts codewords $C^{2:L}$ from the second layer to the L-th layer. Typically, $C^{1}$ primarily encodes semantic information, while $C^{2:L}$ encodes acoustic features such as timbre and prosody~\cite{speechtokenizer}.

The prediction process of the speech token sequence $C^{\text{pred}}$ can be formulated as follows:

\begin{align}
C_t^{1} &= \text{LM}_{\text{AR}}([T, C^{\text{ref}, 1}, C_{0:t-1}^{\text{pred}, 1}]),\\
C_{:}^{l} &= \text{LM}_{\text{NAR}}([T, C^{\text{ref}, 1:l-1}, C_{:}^{\text{pred}, 1:l-1}], l),\quad 2 \le l \le L, \\
C_{t}^{\text{pred}} &= [C_{t}^{1}; C_{t}^{2:L}].
\end{align}

In this section, we introduce our proposed compressed-to-fine language modeling using EnCodec~\cite{encodec} and VALL-E~\cite{valle} as examples. Note that our method is equally applicable to neural audio codecs with a single codebook and decoder-only language models.

\subsection{Short-range Dependency}

In neural codec language models, the prediction of speech tokens is highly dependent on contextual information. To investigate the impact of short-range dependency when predicting the $t$-th speech token $C_t$, we categorize the input token sequence
into the following three types:

$$
[ \underbrace{T, ..., C^{\text{ref}}}_{\text{Prompt Tokens}}, \underbrace{C_{0}, ..., C_{t-N-1}}_{\text{Long-range Tokens}}, \underbrace{C_{t-N}, ..., C_{t-1}}_{\text{Local Tokens}}, \underbrace{C_{t+1}, ..., C_{t+N}}_{\text{Local Tokens}} ]
$$ 

(1) \textit{Prompt tokens}, including text prompt tokens \( T \) and speech prompt tokens \( C^{\text{ref}} \), capture the initial information. 

(2) \textit{Long-range tokens} \( C_{0:t-N-1} \) capture the long-range dependency. 

(3) \textit{Local tokens}, which comprise \( C_{t-N_{\text{AR}}: t-1}^{1} \) for AR decoder \(\text{LM}_{\text{AR}}\) and \( C_{t-N_{\text{NAR}}: t+N_{\text{NAR}}}^{1:l-1} \) for NAR decoder \(\text{LM}_{\text{NAR}}\), capture short-range details.

\noindent where $N$ is the size of the sliding window, $N_{\text{AR}}$ is the size of the causal sliding window for $\text{LM}_{\text{AR}}$,  and $N_{\text{NAR}}$ is the size of the bidirectional sliding window for $\text{LM}_{\text{NAR}}$.

As analyzed in Section~\ref{sec:short_range_dependency}, neural codec language models demonstrate strong short-range dependency: prompt tokens and local tokens already contain the necessary semantic and acoustic information to predict speech tokens, while also maintaining text-speech alignment and paralinguistic features such as timbre. In contrast, long-range tokens may introduce redundant information, which can act as noise and degrade the performance of neural codec language models.

\begin{figure*}[t!]
    \centering
    \includegraphics[scale=0.3]{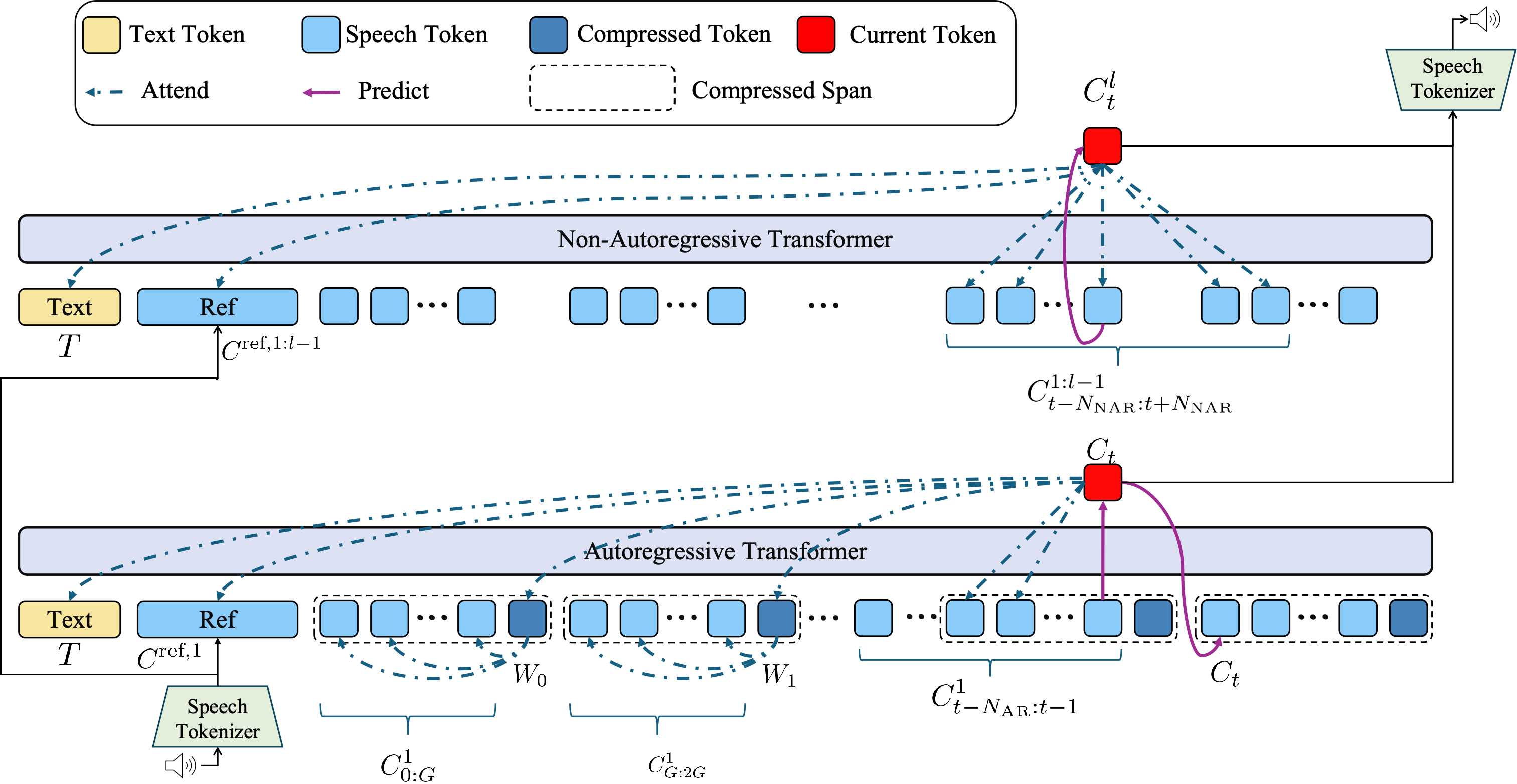}
    \caption{Training process of the proposed compressed-to-fine language modeling for speech generation.}
    \label{valle}
\end{figure*}

Specifically, prompt tokens provide global semantic and acoustic information, while sliding window sizes $N_{\text{AR}}$ and $N_{\text{NAR}}$ constrain the modeling scope of local speech tokens in $\text{LM}_{\text{AR}}$ and $\text{LM}_{\text{NAR}}$. The prompt-local sliding windows, including $[T, C^{\text{ref}, 1}, C_{t-N_{\text{AR}}: t-1}^1]$ for $\text{LM}_{\text{AR}}$ and $[T, C^{\text{ref}, 1:l-1}, C_{t-N_{\text{NAR}}: t+N_{\text{NAR}}}^{1:l-1}]$ for $\text{LM}_{\text{NAR}}$, encode sufficient information for the neural codec language model to predict speech tokens. Compared to using all of speech tokens, relying on prompt-local sliding windows not only improves the quality of generated speech, but also significantly improves inference efficiency. The attention mask of using all tokens, the prompt-local sliding window for $\text{LM}_{\text{AR}}$ and $\text{LM}_{\text{NAR}}$ are depicted in Figure~\ref{fig:attention} (a), (b) and Figure~\ref{fig:nar_attention} (b) in supplementary material.


\subsection{Compressed-to-fine Language Modeling}

Inspired by the analysis of short-range dependency, we propose compressed-to-fine language modeling, which explores \textbf{Fine-grained Initial and Short-range Information} and \textbf{Compressed Long-Range Information}, to optimize modeling of the short-range and long-range speech tokens, respectively. Figure~\ref{valle} demonstrates the detailed approach implemented in VALL-E~\cite{valle} as an example, and Figure~\ref{fig:attention} (c) shows the corresponding attention mask.

\noindent \textbf{Fine-grained Initial and Short-range Information}. To maintain the necessary text-speech alignment and preserve short-term paralinguistic information such as prosody and timbre, only prompt tokens and local tokens are retained. We explicitly maintain initial and local information through prompt-local sliding windows, which attention mask is shown in Figure~\ref{fig:attention}(b), in both autoregressive and non-autoregressive stages:  

\begin{align}
C_{t}^{1} &= \text{LM}_{\text{AR}}([T, C^{\text{ref}}, C_{t-N_{\text{AR}}: t-1}^{1}])\label{eq:ar_sw},\\
C_{t}^{l} &= \text{LM}_{\text{NAR}}([T, C^{\text{ref}}, C_{t-N_{\text{NAR}}: t+N_{\text{NAR}}}^{1:l-1}]), \quad 2 \le l \le L \label{eq:nar_sw}
\end{align}



\noindent \textbf{Compressed Long-range Information}. Although prompt tokens and short-range tokens provide the necessary semantic and acoustic information for speech token prediction, completely discarding long-range tokens may lead to the loss of contextual semantic information. We hypothesize that certain long-range information may offer potential benefits for speech token prediction. Therefore, we compress long-range speech token spans into compact representations, preserving coarse-grained semantic information while filtering out noise and redundancy.

As shown in Figure~\ref{valle}, long-range tokens $C_{0: t-N_{\text{AR}}}^{0}$ are split into $k$ non-overlapping spans, each containing $G$ speech tokens. Obviously, k equals $\lfloor \frac{t-N_{\text{AR}}}{G}\rfloor$. $C_{G \times k: G \times (k+1)}$ denotes the $k$-th span in long-range tokens. $W_k$ is inserted in the last position of every span, and set to only attend to the speech tokens within the $k$-th span by setting attention mask depicted in Figure~\ref{fig:attention}(c), thereby compressing the information of the overall span into a compact representation.

Considering that attending the whole sequence of speech tokens may raise the problem of long context modeling in neural codec language models~\cite{valle2}, language model only attend compact representations $[W_0, ..., W_{k-1}]$ instead of all long-range tokens $C_{0: t-N_{\text{AR}}}^{1}$ when modeling the long-range information. The compact representations $[W_0, ..., W_{k-1}]$, which are compressed from long-range spans $[C_{0:G}^{1}, ..., C_{G \times (k - 1): G \times k}^{1}]$, act as an information bottleneck to filter noise and redundancy while preserving primary semantic information. 

The compression rate $\text{CR} = \frac{\text{frame\_rate}}{G}$ indicates the number of compressed tokens in 1 second of speech in long-range spans.

During parallel training, the speech token sequence $C^{\text{pred}}$ is predicted as follows:

\begin{equation}
\begin{split}
    C^{\text{pred}, 1} = &\text{LM}_{\text{AR}}([T, C^{\text{ref}, 1}, C_{0: G}^{\text{label}, 1}, W_{0}, ..., \\ 
    &C_{G \times (k-1): G \times k}^{\text{label}, 1}, W_{k-1}, C_{t-N_{\text{AR}}: t-1}^{\text{label}, 1}])\label{eq:ar_cf}.
\end{split}
\end{equation}

During inference, the $t$-th token is predicted as follows:

\begin{equation}
\begin{split}
C_{t}^{1} = \text{LM}_{\text{AR}}([T, C^{\text{ref}, 1}, W_{0}, ..., W_{k-1}, C_{t-N_{\text{AR}}: t-1}^{\text{pred}, 1}]) \label{eq:ar_cf_inf}
\end{split}
\end{equation}

Compressed Long-range Information complements the modeling approach of Fine-grained Initial and Short-range Information, while the former provides coarse-grained context, and the latter ensures fine-grained local modeling. Combining these two aspects in the proposed Compressed-to-fine Language Modeling improves speech token generation.

\subsection{Training and Inference}

\noindent \textbf{Training.} The autoregressive decoder $\text{LM}_{\text{AR}}$ adopts the complete compressed-to-fine language modeling with both Fine-grained Initial and Short-range Information and Compressed Long-range Information, and $\text{LM}_{\text{NAR}}$ only employs the prompt-local sliding window for Fine-grained Initial and Short-range Information. The training process is depicted as Eq. \ref{eq:ar_cf} for $\text{LM}_{\text{AR}}$, and Eq. \ref{eq:nar_sw} for $\text{LM}_{\text{NAR}}$.

$W$ is only used to compress the spans of speech tokens, and is not predicted by the neural codec language model. The language model is trained with the cross-entropy loss between the ground truth $C^{\text{label}}$ and $C^{\text{pred}}$, while the loss at the positions of $W$ is omitted.


\noindent \textbf{Inference.} The inference process is described by Eq.~\ref{eq:ar_cf_inf} for $\text{LM}_{\text{AR}}$ and Eq.~\ref{eq:nar_sw} for $\text{LM}_{\text{NAR}}$. During the prediction of the $t$-th speech token, $C_{0:t-N_{\text{AR}}}^{1}$ has already been compressed into $[W_0, ..., W_k]$. We investigate two inference strategies for autoregressive decoding: 

(1) \textbf{Vanilla Inference}: In this strategy, within the attention mask, $[T, C^{\text{ref}}]$ and $[W_0, ..., W_{k-1}]$ are kept visible, while $C_{0:t-N_{\text{AR}}}^{1}$ remains invisible. After every $G$ speech tokens are predicted, the next input token is the compressed token $W$, which means that the information of the current span is compressed into $W$ in this inference step. Thus, the KV cache for this step is preserved, but the output predicted token is meaningless and will be ignored in the following inference steps. The vanilla inference strategy is compatible with existing architectures (e.g., decoder-only transformer and VALL-E), but it does not improve inference speed.

(2) \textbf{Faster inference}: When predicting the $t$ speech token, this strategy allows the KV cache of $C_{0:t-N_{\text{AR}}}^{1}$ to be evicted while retaining the KV cache of $[T, C^{\text{ref}}]$ and $[W_0, ..., W_{k-1}]$, thus reducing the computational complexity from $O(N_p+T)$ to $O(N_p+\frac{T}{G}+N_{\text{AR}})$ and accelerating the inference speed. Here, $N_p$ denotes the number of prompt tokens $[T, C^{\text{ref}}]$. The pseudo code is depicted in Algorithm~\ref{inference} in the appendix in the supplementary material. 

\textbf{Compatibility with mainstream transformer-based architecture.} The proposed compressed-to-fine language modeling does not depend on specific neural audio codecs and neural codec language models. Instead, \textbf{it is compatible with existing transformer-based architectures}, including decoder-only transformer, hybrid transformers such as VALL-E, and even text-based LLMs, by inserting the compressed token $W$ and setting the appropriate attention mask during both training and inference processes.

\section{Experimental Setup}
\begin{table}[h]
    \centering
    \caption{Details of compressed-to-fine language modeling across various neural audio codecs and neural codec language models. The symbol "-" indicates that $N_{\text{NAR}}$ is not used.}
    \scalebox{0.78}{
    \begin{tabular}{ccc|c|ccc|c}
    \toprule
        \multirow{2}{*}{\makecell[c]{Neural\\Audio Codec}} & \multirow{2}{*}{\makecell[c]{sampling\\rate}} & \multirow{2}{*}{\makecell[c]{frame\\rate}} & \multirow{2}{*}{\makecell[c]{Neural Codec\\ Language Model}} & \multicolumn{3}{c|}{$\text{LM}_{\text{AR}}$} & $\text{LM}_{\text{NAR}}$ \\ \cline{5-8}
         &  &  & & $G$ & CR & $N_{\text{AR}}$ & $N_{\text{NAR}}$ \\ \midrule
        HuBERT & 16000 & 50Hz & decoder-only & 10 & 5 & 50 & - \\ \midrule
        BigCodec & 16000 & 80Hz & decoder-only & 16 & 5 & 80 & - \\ \midrule
        WavTokenizer & 24000 & 75Hz & decoder-only & 15 & 5 & 75 & - \\ \midrule
        EnCodec & 16000 & 50Hz & VALL-E & 10 & 5 & 50 & 50 \\ \bottomrule
    \end{tabular}}
    \label{tab:cf_setup}
\end{table}

\begin{table*}[!t]
    \centering
    \caption{\textbf{Objective} speech generation results on LibriSpeech test set. \textbf{Bold} means the best result, and \underline{underline} means the second-best result. \textit{VALL-E$^*$} denotes \textit{the improved implementation of VALL-E}. The superscript \textit{CF} denotes \textit{compressed-to-fine language modeling}, $\dagger$ denotes results cited from the original paper, and $\diamondsuit$ means the results are inferred from the official checkpoints. The rest results are from model implementation, training and evaluation in this work. Scale refers to the model size or training data scale.}
    \scalebox{0.95}{
    \begin{tabular}{cc|lc|ccc}
    \toprule
        \multirow{2}{*}{\makecell[c]{Neural\\Audio Codec}} & \multirow{2}{*}{Bandwidth} & \multirow{2}{*}{\makecell[c]{Neural Codec\\ Language Model}} & \multirow{2}{*}{Scale} & \multicolumn{3}{c}{\textbf{Objective Metrics}} \\
         & & & & WER$\downarrow$ & SIM$\uparrow$ & UTMOS$\uparrow$ \\ \midrule
        Ground Truth & - & - & - & - & 69.72\% & 4.15 \\
    \midrule
    \addlinespace[-0.5pt] 
    \midrule
        \multirow{2}{*}{EnCodec~\cite{encodec}} & \multirow{2}{*}{2.2 kbps} & \multirow{2}{*}{VoiceCraft$^\diamondsuit$~\cite{voicecraft}} & 330M,9Kh & 4.46 & 52.15\% & 3.55 \\
         &  &  & 830M,9Kh & 3.31 & 56.75\% & 3.78 \\
    \midrule
        Mel VQ-VAE & - & XTTS\_v2$^\diamondsuit$~\cite{xtts} & 27Kh & 2.98 & \underline{59.14\%} & 3.94 \\
    \midrule
        \multirow{1}{*}{SpeechTokenizer~\cite{speechtokenizer}} & \multirow{1}{*}{4.0 kbps} & USLM$^\diamondsuit$~\cite{speechtokenizer} & 960h & 6.25 & 56.93\% & 3.12 \\
    \midrule
        \multirow{2}{*}{EnCodec~\cite{encodec}} & \multirow{2}{*}{6.0 kbps} & VALL-E$^{\dagger}$~\cite{valle} & 60Kh & 5.90 & 58.00\% & - \\
        &  & VALL-E R$^{\dagger}$~\cite{valler} & 960h & 3.18 & - & - \\
    \midrule
    \addlinespace[-0.5pt] 
    \midrule
        \multirow{2}{*}{HuBERT~\cite{hubert}} & \multirow{2}{*}{0.5 kbps} & decoder-only & \multirow{2}{*}{960h} & 7.04 & - & 4.07 \\ 
        &  & decoder-only$^{\text{CF}}$ &  & 5.86 & - & 4.08 \\
    \midrule
        \multirow{2}{*}{BigCodec~\cite{bigcodec}} & \multirow{2}{*}{1.04 kbps} & decoder-only & \multirow{2}{*}{960h} & 16.58 & 46.04\% & 3.78 \\ 
        &  & decoder-only$^{\text{CF}}$ &  & 10.90 & 46.51\% & 4.01 \\
    \midrule
        \multirow{2}{*}{WavTokenizer~\cite{wavtokenizer}} & \multirow{2}{*}{0.9 kbps} & decoder-only & \multirow{2}{*}{960h} & 5.94 & 47.59\% & 4.14 \\ 
        &  & decoder-only$^{\text{CF}}$ &  & 3.40 & 47.99\% & \textbf{4.19} \\
    \midrule
        \multirow{3}{*}{EnCodec~\cite{funcodec}} & \multirow{3}{*}{4.0 kbps} & VALL-E$^{\text{vanilla}}$ & \multirow{3}{*}{960h} & 7.59 & 56.08\% & 4.13 \\ 
        & & VALL-E$^*$ & & 5.23 & 56.64\% & 4.16 \\ 
        &  & VALL-E$^{*\text{CF}}$ & & \underline{2.36} & \textbf{59.75\%} & \underline{4.18} \\ \midrule
        \multirow{2}{*}{EnCodec~\cite{funcodec}} & \multirow{2}{*}{4.0 kbps} & VALL-E$^*$ & \multirow{2}{*}{20Kh} & 5.05 & 54.83\% & 4.09 \\ 
        &  & VALL-E$^{*\text{CF}}$ & & \textbf{2.21} & 57.72\% & 4.15 \\
    \bottomrule
    \end{tabular}
    }
    \label{tab:tts}
\end{table*}



\noindent \textbf{Baseline Models and Implementation Details.} We evaluate the proposed method on speech generation performance on the test set of LibriSpeech~\cite{librispeech}. As part of the baselines, we cite results from representative neural codec language models on this test set, including VoiceCraft~\cite{voicecraft}, SpeechTokenizer-based USLM~\cite{speechtokenizer} and XTTS v2~\cite{xtts}.


To evaluate the effectiveness of the proposed method, we also systematically evaluate the proposed method across different \textit{representative and competitive} neural audio codecs and neural codec language models.  Neural audio codecs can be broadly categorized into multi-codebook and single-codebook codecs, with speech tokens from these codecs being predicted by VALL-E and decoder-only transformer-based language models. We conduct evaluations on the following multi-codebook and single-codebook codecs.

\noindent \textit{Multi-codebook Codecs.} We train EnCodec~\cite{encodec} (8 codebooks, each with 1024 codewords, reproduced by FunCodec~\cite{funcodec}) on LibriTTS~\cite{libritts} and VALL-E~\cite{valle} on LibriSpeech~\cite{librispeech}. EnCodec is optimized by the Adam optimizer~\cite{2014adam} with an initial learning rate of 3e-4. Audio samples are truncated to 1.28 seconds, resampled to 16 kHz, with a batch size of 384. We optimize VALL-E by the AdamW optimizer with an initial learning rate of 0.01~\cite{xcodec} and beta parameters (0.5, 0.9).

\noindent \textit{Single-codebook Codecs.} We use pre-trained HuBERT~\cite{hubert} and its corresponding unit-based HiFiGAN~\cite{hifigan}, WavTokenizer~\cite{wavtokenizer}, and BigCodec~\cite{bigcodec} as speech tokenizers. The decoder-only transformer is adopted as the neural codec language model, and trained on LibriSpeech, using the AdamW optimizer with an initial learning rate of 9.5e-5 and beta parameters (0.5, 0.9).

All transformer models in both VALL-E and decoder-only architectures for neural codec language models have a dimension of 1024, 12 transformer blocks, and 16 attention heads. To improve performance, we incorporate several components of the Llama model~\cite{2023llama} into the transformer block, including RoPE and the SiLU activation function. We denote 
the improved version of VALL-E by \textbf{VALL-E}$^*$ in Table ~\ref{tab:tts}.


~\cite{guo2025recent, cho2024sylber, syllablelm} suggest that 5Hz discrete units can effectively align with syllables, capturing essential linguistic structure. Thus, we set the compression rate $\text{CR} = 5$ to compress the speech tokens of a 1-second span into 5 compact representations, retaining coarse-grained semantic information. Hyperparameters of compressed-to-fine language modeling are detailed in Table~\ref{tab:cf_setup}.
We train all VALL-E$^*$ and transformer-decoder-only neural codec language models using 960 hours of LibriSpeech training data. To assess data scaling, we expand training data to 44,000 hours by adding MLS~\cite{MLS} dataset.


\noindent \textbf{Evaluation Metrics.}

\noindent \textit{Objective evaluation.} Whisper~\cite{whisper} is used to transcribe the generated speech and then calculate Word Error Rate (WER). To evaluate speaker similarity (SIM), we use 3D-speaker toolkit~\cite{3dspeaker} to extract speaker embeddings from the generated speech and reference speech, and compute the cosine similarity between the normalized embeddings. UTMOS~\cite{utmos} also serves as an automatic Mean Opinion Score (MOS) prediction system to assess the naturalness of the generated speech. 

RTF (Real-Time Factor) is computed as the ratio of the time required to generate speech to the duration of the generated speech. A lower RTF value indicates faster speech generation.

\noindent \textit{Subjective evaluation.} We randomly select speech samples from LibriSpeech~\cite{librispeech} test set to conduct MOS~\cite{chu2006objective}, Similarity Mean Opinion Score (SMOS)~\citep{chu2006objective}, and Prosody MOS (PMOS) evaluations. MOS assesses naturalness of the generated speech, SMOS measures timbre similarity between the generated speech and voice of the original speaker, and PMOS evaluates prosody of the generated speech.

\section{Results and Analysis}

\subsection{Speech Generation Results}

\textbf{Objective Evaluation}. Table~\ref{tab:tts} presents objective speech generation results of various neural codec language models on LibriSpeech test set. Since VALL-E~\cite{valle} and VALL-E R~\cite{valler} do not provide reproducible code, we cite the results from their original papers. Also, since some models they use to evaluate SIM scores differ significantly from other baselines, we only cite their SIM results when the comparison is fair. According to Table ~\ref{tab:tts}, we draw the following conclusions.

(1) For both single-codebook codecs (e.g., HuBERT, WavTokenizer, BigCodec) and multi-codebook codecs (e.g., EnCodec), and various neural codec language models (e.g., VALL-E, decoder-only transformer),  applying compressed-to-fine language modeling consistently and substantially improves generation accuracy (WER), speaker similarity (SIM), and speech quality (UTMOS), demonstrating the effectiveness of the proposed method in enhancing speech intelligibility and naturalness. Our method improves VALL-E (20kh) by 2.84\% absolute and 56.24\% relative reduction in WER and 2.89\% absolute and 5.27\% relative improvement in speaker similarity. When using WavTokenizer as speech tokenizer and decoder-only language model, applying the proposed method also achieves 2.54\% absolute and 42.76\% relative reduction in WER.


(2) Prior studies~\cite{speechtokenizer} found that in VALL-E~\cite{valle}, \(\text{LM}_{\text{AR}}\) primarily handles semantic information, while \(\text{LM}_{\text{NAR}}\) processes acoustic information. Applying compressed-to-fine language modeling on the improved VALL-E$^*$ yields substantial gains, especially in WER and SIM. These results suggest that the proposed method optimizes both \(\text{LM}_{\text{AR}}\) and \(\text{LM}_{\text{NAR}}\), improving semantic modeling (measured by WER) and acoustic modeling (measured by SIM, UTMOS).


\noindent \textbf{Subjective Evaluation}. As shown in Table~\ref{tab:tts_subjective}, across diverse neural audio codecs and language models, using compressed-to-fine language modeling generally improves subjective scores, highlighting its effectiveness in enhancing the quality of generated speech.


\begin{table}[!t]
    \centering
    \caption{\textbf{Subjective} speech generation performance on LibriSpeech test set.}
    \scalebox{0.86}{
    \begin{tabular}{c|l|ccc}
    \toprule
        \multirow{2}{*}{\makecell[c]{Neural\\Audio Codec}} & \multirow{2}{*}{\makecell[c]{Neural Codec\\ Language Model}} & \multicolumn{3}{c}{Subjective Metrics} \\
         & & MOS$\uparrow$ & SMOS$\uparrow$ & PMOS$\uparrow$ \\ \midrule
        Ground Truth & - & 4.04±0.10 & 4.21±0.19 & 4.12±0.21 \\
    \midrule
        \multirow{2}{*}{HuBERT~\cite{hubert}} & decoder-only & 2.96±0.10 & - & 3.08±0.08 \\
        &  decoder-only$^{\text{CF}}$ & 3.10±0.12 & - & 3.21±0.06 \\
    \midrule
        \multirow{2}{*}{WavTokenizer~\cite{wavtokenizer}} & decoder-only & 3.58±0.12 & 3.71±0.06 & 3.54±0.10 \\
        &  decoder-only$^{\text{CF}}$ & 3.62±0.26 & 3.79±0.06 & 3.60±0.12 \\
    \midrule
        \multirow{2}{*}{EnCodec~\cite{funcodec}} & VALL-E & 3.29±0.27 & 3.79±0.19 & 3.12±0.17 \\
        & VALL-E$^{\text{CF}}$ & 3.48±0.28 & 4.08±0.20 & 3.75±0.31 \\
    \bottomrule
    \end{tabular}
    }
    \label{tab:tts_subjective}
\end{table}

\subsection{Analysis of Short-range Dependency}
\label{sec:short_range_dependency}

\begin{table}[!h]
    \centering
    \caption{Performance on the LibriSpeech test set from neural codec language models trained with various prompt-local sliding window sizes. \( N_{\text{AR}} \) and \( N_{\text{NAR}} \) are the sliding window size for \( \text{LM}_{\text{AR}} \) and \( \text{LM}_{\text{NAR}} \), respectively. When $N_{\text{AR}}$ is employed, the input token sequence for $\text{LM}_{\text{AR}}$ is $[T, C^{\text{ref}, 1}, C_{t-N_{\text{AR}}: t-1}^{1}]$. When $N_{\text{NAR}}$ is employed, the input token sequence for $\text{LM}_{\text{NAR}}$ is $[T, C^{\text{ref}, 1:L}, C_{t-N_{\text{NAR}}: t+N_{\text{NAR}}}^{1:L}]$. \textbf{-} indicates that sliding window is disabled.}
    \scalebox{0.9}{
    \begin{tabular}{cc|c|c|ccc}
    \toprule
        \multirow{2}{*}{\makecell[c]{Neural\\ Codec}} & \multirow{2}{*}{\makecell[c]{Language\\Model}} & \multirow{2}{*}{$N_{\text{AR}}$} & \multirow{2}{*}{$N_{\text{NAR}}$} & \multicolumn{3}{c}{Objective} \\ 
        & & & & WER↓ & SIM↑ & UTMOS↑ \\ \midrule
        \multirow{8}{*}{EnCodec} & \multirow{8}{*}{VALL-E} & - & - & 5.23 & 56.64\% & 4.16 \\ \cmidrule{3-7}
        & & 30 & - & 4.95 & 56.80\% & 4.16 \\
        & & 50 & - & \textbf{4.07} & \textbf{57.42\%} & 4.14 \\
        & & 120 & - & 4.29 & 57.35\% & 4.17 \\
        & & 200 & - & 4.48 & 56.59\% & \textbf{4.20} \\ \cmidrule{3-7}
        & & - & 30 & 3.95 & 58.40\% & 4.14 \\
        & & - & 50 & \textbf{3.24} & \textbf{58.65\%} & 4.12 \\
        & & - & 120 & 3.53 & 57.87\% & \textbf{4.17} \\ \midrule
        \multirow{3}{*}{\makecell[c]{Wav\\Tokenizer}} & \multirow{3}{*}{\makecell[c]{decoder\\-only}} & - & - & 5.94 & 47.59 & 4.14 \\ \cmidrule{3-7}
        & & 75 & - & 7.85 & 47.63 & 4.11 \\
        & & 150 & - & \textbf{4.93} & \textbf{47.72} & \textbf{4.15} \\ \bottomrule
    \end{tabular}}
    \label{tab:short_range_in_valle}
\end{table}



Previous research~\cite{speechtokenizer} found that in VALL-E~\cite{valle}, \(\text{LM}_{\text{AR}}\) primarily processes encoding and decoding of semantic information, while \(\text{LM}_{\text{NAR}}\) handles the acoustic information, including timbre and prosody. Thus, by adjusting \( C_{t-N_{\text{AR}}: t-1}^{1} \) and \( C_{t-N_{\text{NAR}}: t+N_{\text{NAR}}}^{1:l-1} \) depicted respectively in Figure~\ref{fig:attention}(b) and Figure~\ref{fig:nar_attention}(b) in the appendix in the supplementary material, we can observe how local semantic and acoustic tokens affect speech token prediction. Table~\ref{tab:short_range_in_valle} shows the results of training VALL-E and a decoder-only transformer with different prompt-local sliding window sizes. We make the following findings.

(1) In \(\text{LM}_{\text{AR}}\), relying solely on prompt tokens and local tokens $[T, C^{\text{ref, 1}}, C_{t-N_{\text{AR}}:t}^{1}]$ substantially reduces WER, indicating that \(\text{LM}_{\text{AR}}\) primarily relies on local tokens to predict semantically rich codewords.
(2) In \(\text{LM}_{\text{NAR}}\), relying solely on prompt and local tokens $[T, \\C^{ref, 1:l-1}, C_{t-N_{\text{NAR}}: t+N_{\text{NAR}}}^{1:l-1}]$ notably enhances SIM, suggesting that \(\text{LM}_{\text{NAR}}\) primarily depends on local tokens to predict acoustically rich codewords.
(3) As the sliding window size increases, model performance gradually improves, but once \( N_{\text{AR}} \) or \( N_{\text{NAR}} \) surpasses a certain threshold, performance gains plateau. This may be because long-range tokens contain redundant information, and an excessively large sliding window might introduce more redundancies, undermining robustness of the model.

In summary, prompt tokens and local tokens already encompass the necessary semantic and acoustic information, effectively supporting the prediction of speech tokens in neural codec language models. In contrast, redundant information in long-range tokens might introduce noise and redundancy, reducing robustness of the model and degrading its performance.

\begin{table*}[!ht]
    \centering
    \caption{Ablation study on the LibriSpeech test set. The symbol "-" indicates that $G$ or $N$ is not used.}
    \scalebox{0.85}{
    \begin{tabular}{ccc|c|ccc|c|ccc}
    \toprule
        \multirow{2}{*}{\makecell[c]{Neural\\Audio Codec}} & \multirow{2}{*}{\makecell[c]{sampling\\rate}} & \multirow{2}{*}{\makecell[c]{frame\\rate}} & \multirow{2}{*}{\makecell[c]{Neural Codec\\ Language Model}} & \multicolumn{3}{c|}{$\text{LM}_{\text{AR}}$} & $\text{LM}_{\text{NAR}}$ & \multicolumn{3}{c}{Objective Metrics}  \\ \cmidrule{5-11}
         &  &  & & $G$ & CR & $N_{\text{AR}}$ & $N_{\text{NAR}}$ & WER↓ & SIM↑ & UTMOS↑ \\ \midrule
        \multirow{5}{*}{WavTokenizer} & \multirow{5}{*}{24000} & \multirow{5}{*}{75Hz} & \multirow{5}{*}{decoder-only} & - & - & - & - & 5.94 & 47.59\% & 4.14 \\ \cmidrule{5-11}
         &  &  &  & - & - & 75 & - & 7.85 & 47.63\% & 4.11 \\ \cmidrule{5-11}
         &  &  &  & - & - & 150 & - & 4.96 & \textbf{48.04\%} & 4.16 \\ \cmidrule{5-11}
         &  &  &  & 15 & 5Hz & 75 & - & \textbf{3.40} & 47.99\% & \textbf{4.19} \\
         &  &  &  & 15 & 5Hz & 150 & - & 4.13 & 47.97\% & 4.18 \\
         \midrule
        \multirow{10}{*}{EnCodec} & \multirow{10}{*}{16000} & \multirow{10}{*}{50Hz} & \multirow{10}{*}{VALL-E} & - & - & - & - & 5.23 & 56.64\% & 4.16 \\ \cmidrule{5-11}
         & & & & 5 & 10Hz & 50 & - & 3.30 & 57.98\% & \textbf{4.21} \\
         &  &  &  & 10 & 5Hz & 50 & - & 2.87 & 57.76\% & \textbf{4.21} \\
         &  &  &  & 15 & 3.33Hz & 50 & - & 3.84 & 56.62\% & 4.19 \\ \cmidrule{5-11}
         &  &  &  & 10 & 5Hz & 30 & - & 2.95 & 57.41\% & 4.20 \\
         &  &  &  & 10 & 5Hz & 50 & - & 2.87 & 57.76\% & \textbf{4.21} \\
         &  &  &  & 10 & 5Hz & 120 & - & 4.66 & 57.27\% & 4.16 \\ \cmidrule{5-11}
         &  &  &  & 10 & 5Hz & 50 & 30 & 3.02 & 59.93\% & 4.18 \\
         &  &  &  & 10 & 5Hz & 50 & 50 & \textbf{2.36} & \textbf{59.75\%} & 4.18 \\
         &  &  &  & 10 & 5Hz & 50 & 120 & 3.00 & 58.49\% & 4.19 \\ \bottomrule
    \end{tabular}}
    \label{tab:ablation}
\end{table*}

\subsection{Visualization of Attention Weights} 


\begin{figure}[h]
    \centering 
    \begin{minipage}[t]{\columnwidth}
        \centering  
        \includegraphics[scale=0.39]{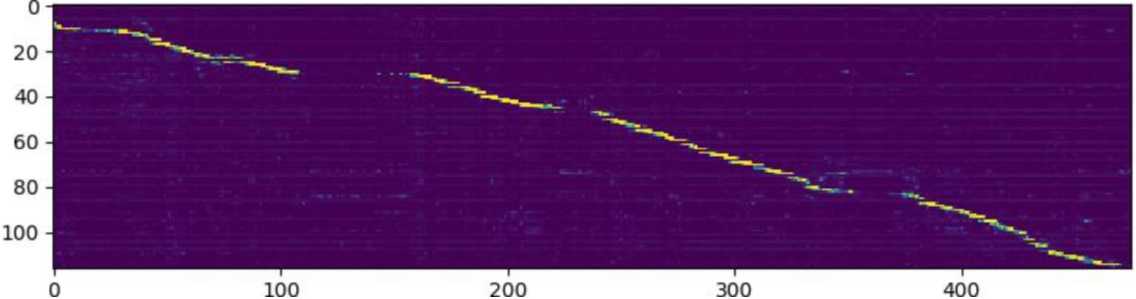}
        \caption*{Fig 3 (a): Attention weight of VALL-E}  
        \label{fig:valle_attention_score}
    \end{minipage}
    \vspace{\baselineskip} 
    \begin{minipage}[t]{\columnwidth}
        \centering
        \includegraphics[scale=0.37]{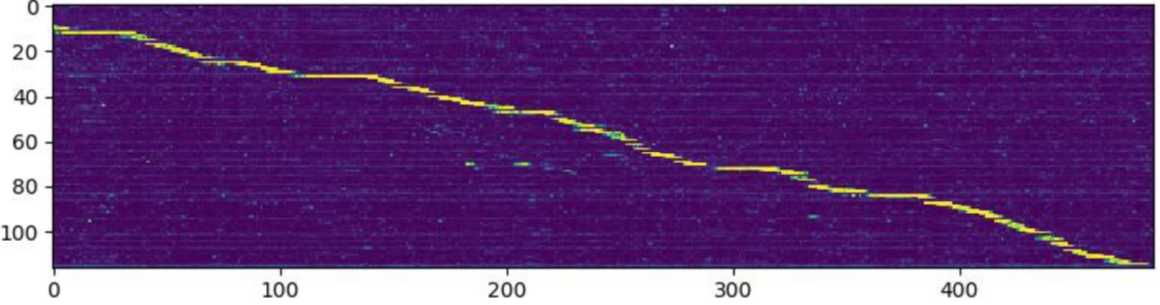}
        \caption*{Fig 3 (b): Attention weight of VALL-E with the prompt-local sliding window.}
        \label{fig:valle_sliding_attention_score}
    \end{minipage}
    \vspace{\baselineskip} 
    \begin{minipage}[t]{\columnwidth}
        \centering
        \includegraphics[scale=0.395]{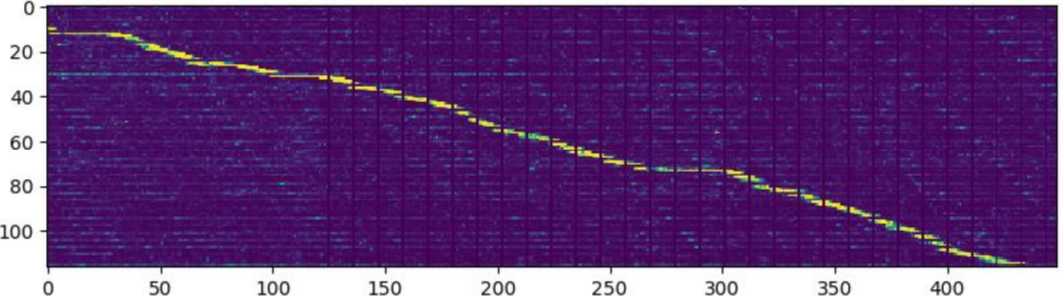}
        \caption*{Fig 3 (c): Attention weight of VALL-E with compressed-to-fine language modeling. The vertical black line indicates the compressed token.}
        \label{fig:valle_cf_attention_score}
    \end{minipage}
    \caption{Visualization of attention weight between text prompt tokens and predicted speech tokens. This example is from LibriSpeech test set. The brighter the color, the greater the attention weight, and the tighter the alignment between text and speech tokens. 
    }
    \label{fig:attention_score}
\end{figure}

To understand the mechanism behind the compressed-to-fine language modeling, we conduct visualization analysis. Figure~\ref{fig:attention_score} shows the average attention weights between text prompt tokens $T$ and predicted speech tokens $C^{\text{pred}}$ across all attention heads during autoregressive inference.
(1) As shown in Figure~\ref{fig:attention_score} (a), the text-speech alignment path is discontinuous in VALL-E, indicating occasional misalignment during speech generation, which may lead to misreading, omission, or repetition. (2) As shown in Figure~\ref{fig:attention_score} (b) and (c), with the prompt-local sliding window and compressed-to-fine language modeling, the alignment path becomes continuous, monotonic, and brighter, ensuring all speech tokens align with text prompt tokens, satisfying locality, monotonicity, and completeness as defined in ~\cite{valler}. (3) By introducing the prompt-local sliding window, the speech token sequence $[T, C^{\text{ref}, 1}, C_{0: t-1}^{1}]$ is reduced to $[T, C^{\text{ref}, 1}, C_{t-N_{\text{AR}}: t-1}^{1}]$, significantly shortening speech token sequence length and making it easier for the language model to capture the alignment relationship between text and speech tokens. The compressed-to-fine language modeling demonstrates similar alignment path, verifying that compressed tokens do not disrupt alignment between text and speech, and that prompt-local sliding window continues its critical role in enhancing this alignment.


\subsection{Accelerate Inference}
We also compare inference efficiency in Table~\ref{tab:rtf}. The results demonstrate that compressed-to-fine language modeling accelerates the inference process and achieves a 20\% relative improvement in inference speed at a high frame rate of 75.

\begin{table}[!ht]
    \setlength{\extrarowheight}{2pt}
    \centering
    \caption{Inference speed of autoregressive (AR) decoding for speech generation on LibriSpeech test set.}
    \scalebox{0.85}{
    \begin{tabular}{cc|l|c|c}
    \toprule
        \multirow{2}{*}{\makecell[c]{Neural\\Audio Codec}} & \multirow{2}{*}{\makecell[c]{Frame\\Rate}} & \multirow{2}{*}{\makecell[c]{Neural Codec\\Language Model}} & \multirow{2}{*}{\makecell[c]{Average Time \\Per AR Step (/s)↓}} & \multirow{2}{*}{RTF↓} \\ 
         & & & & \\ \midrule
        \multirow{2}{*}{HuBERT} & \multirow{2}{*}{50} & Decoder-Only & 0.0102 & 0.7687 \\ \cmidrule{3-5}
        ~ & ~ & $\text{Decoder-Only}^{\text{CF}}$ & 0.0085 & 0.6437 \\ \midrule
        \multirow{2}{*}{WavTokenizer} & \multirow{2}{*}{75} & Decoder-Only & 0.0105 & 0.7911 \\ \cmidrule{3-5}
        ~ & ~ & $\text{Decoder-Only}^{\text{CF}}$ & 0.0086 & 0.6482 \\ \bottomrule
    \end{tabular}}
    \label{tab:rtf}
\end{table}

\subsection{Ablation Study}

Table~\ref{tab:short_range_in_valle} shows the results using only the prompt-local sliding window size $N$, suggesting that this simple technique significantly improves intelligibility and speaker similarity of generated speech. Table~\ref{tab:ablation} shows the performance of neural codec language models under different $G$ and $N$ when applying \textit{compressed-to-fine language modeling}. $G$ denotes the size of the compressed long-range span, i.e., compressing $G$ consecutive speech tokens into 1 compact representation. We make the following findings.


(1) \textbf{Impact of $G$}: In VALL-E, as $G$ increases, model performance first decreases, then increases, and finally decreases again. When $G$ is too small (e.g., 5), redundant information is not effectively filtered; when $G$ is too large (e.g., 15), excessive information is lost. Both scenarios hinder the language model’s ability to predict speech tokens, indicating an optimal value for $G$.

(2) \textbf{Impact of $N$}: Similarly, as $N$ increases, model performance also exhibits a trend of first decreasing, then increasing, and finally decreasing again. 
When $N$ is too small (e.g., 30), insufficient information leads to degraded model performance; when $N$ is too large (e.g., 120), redundant information is introduced and similarly affects performance, also suggesting an optimal value for $N$.

(3) We find the \textbf{optimal parameter setting} is compressing 1 second of speech tokens into 5 compact representations in the long-range span and retaining neighboring 1-second speech tokens in the short-range span. This observation aligns with previous research: speech typically contains about 5 syllables per second~\cite{guo2025recent, cho2024sylber, syllablelm}, thus it is reasonable to compress 50 (for EnCodec) or 75 (for WavTokenizer) speech tokens per second by a factor of 5 and lead to 10 (for EnCodec) or 15 (for WavTokenizer) compact representations.

\section{Conclusion}

We introduce compressed-to-fine language modeling to alleviate sparsity and redundancy in neural codec language models. Our method divides speech tokens into two components: (1) fine-grained initial and local information. (2) compressed long-range information.
Experiments demonstrate the effectiveness and generalizability of our approach. Future work will explore dynamic adaptation of compression and window sizes in diverse speech understanding and generation tasks.

\clearpage

\bibliographystyle{ACM-Reference-Format}
\bibliography{sample-base}


\begin{thebibliography}{56}


\ifx \showCODEN    \undefined \def \showCODEN     #1{\unskip}     \fi
\ifx \showISBNx    \undefined \def \showISBNx     #1{\unskip}     \fi
\ifx \showISBNxiii \undefined \def \showISBNxiii  #1{\unskip}     \fi
\ifx \showISSN     \undefined \def \showISSN      #1{\unskip}     \fi
\ifx \showLCCN     \undefined \def \showLCCN      #1{\unskip}     \fi
\ifx \shownote     \undefined \def \shownote      #1{#1}          \fi
\ifx \showarticletitle \undefined \def \showarticletitle #1{#1}   \fi
\ifx \showURL      \undefined \def \showURL       {\relax}        \fi
\providecommand\bibfield[2]{#2}
\providecommand\bibinfo[2]{#2}
\providecommand\natexlab[1]{#1}
\providecommand\showeprint[2][]{arXiv:#2}

\bibitem[An et~al\mbox{.}(2024)]%
        {funaudiollm}
\bibfield{author}{\bibinfo{person}{Keyu An}, \bibinfo{person}{Qian Chen}, \bibinfo{person}{Chong Deng}, \bibinfo{person}{Zhihao Du}, \bibinfo{person}{Changfeng Gao}, \bibinfo{person}{Zhifu Gao}, \bibinfo{person}{Yue Gu}, \bibinfo{person}{Ting He}, \bibinfo{person}{Hangrui Hu}, \bibinfo{person}{Kai Hu}, {et~al\mbox{.}}} \bibinfo{year}{2024}\natexlab{}.
\newblock \showarticletitle{Funaudiollm: Voice understanding and generation foundation models for natural interaction between humans and llms}.
\newblock \bibinfo{journal}{\emph{arXiv preprint arXiv:2407.04051}} (\bibinfo{year}{2024}).
\newblock


\bibitem[Aylett and Turk(2006)]%
        {aylett2006language}
\bibfield{author}{\bibinfo{person}{Matthew Aylett} {and} \bibinfo{person}{Alice Turk}.} \bibinfo{year}{2006}\natexlab{}.
\newblock \showarticletitle{Language redundancy predicts syllabic duration and the spectral characteristics of vocalic syllable nuclei}.
\newblock \bibinfo{journal}{\emph{The Journal of the Acoustical Society of America}} \bibinfo{volume}{119}, \bibinfo{number}{5} (\bibinfo{year}{2006}), \bibinfo{pages}{3048--3058}.
\newblock


\bibitem[Baade et~al\mbox{.}(2024)]%
        {syllablelm}
\bibfield{author}{\bibinfo{person}{Alan Baade}, \bibinfo{person}{Puyuan Peng}, {and} \bibinfo{person}{David Harwath}.} \bibinfo{year}{2024}\natexlab{}.
\newblock \showarticletitle{Syllablelm: Learning coarse semantic units for speech language models}.
\newblock \bibinfo{journal}{\emph{arXiv preprint arXiv:2410.04029}} (\bibinfo{year}{2024}).
\newblock


\bibitem[Baas et~al\mbox{.}(2025)]%
        {mars6}
\bibfield{author}{\bibinfo{person}{Matthew Baas}, \bibinfo{person}{Pieter Scholtz}, \bibinfo{person}{Arnav Mehta}, \bibinfo{person}{Elliott Dyson}, \bibinfo{person}{Akshat Prakash}, {and} \bibinfo{person}{Herman Kamper}.} \bibinfo{year}{2025}\natexlab{}.
\newblock \showarticletitle{MARS6: A Small and Robust Hierarchical-Codec Text-to-Speech Model}.
\newblock \bibinfo{journal}{\emph{arXiv preprint arXiv:2501.05787}} (\bibinfo{year}{2025}).
\newblock


\bibitem[Borsos et~al\mbox{.}(2023)]%
        {soundstorm}
\bibfield{author}{\bibinfo{person}{Zal{\'a}n Borsos}, \bibinfo{person}{Matt Sharifi}, \bibinfo{person}{Damien Vincent}, \bibinfo{person}{Eugene Kharitonov}, \bibinfo{person}{Neil Zeghidour}, {and} \bibinfo{person}{Marco Tagliasacchi}.} \bibinfo{year}{2023}\natexlab{}.
\newblock \showarticletitle{Soundstorm: Efficient parallel audio generation}.
\newblock \bibinfo{journal}{\emph{arXiv preprint arXiv:2305.09636}} (\bibinfo{year}{2023}).
\newblock


\bibitem[Casanova et~al\mbox{.}(2024)]%
        {xtts}
\bibfield{author}{\bibinfo{person}{Edresson Casanova}, \bibinfo{person}{Kelly Davis}, \bibinfo{person}{Eren G{\"o}lge}, \bibinfo{person}{G{\"o}rkem G{\"o}knar}, \bibinfo{person}{Iulian Gulea}, \bibinfo{person}{Logan Hart}, \bibinfo{person}{Aya Aljafari}, \bibinfo{person}{Joshua Meyer}, \bibinfo{person}{Reuben Morais}, \bibinfo{person}{Samuel Olayemi}, {et~al\mbox{.}}} \bibinfo{year}{2024}\natexlab{}.
\newblock \showarticletitle{Xtts: a massively multilingual zero-shot text-to-speech model}.
\newblock \bibinfo{journal}{\emph{arXiv preprint arXiv:2406.04904}} (\bibinfo{year}{2024}).
\newblock


\bibitem[Chen et~al\mbox{.}(2025a)]%
        {minmo}
\bibfield{author}{\bibinfo{person}{Qian Chen}, \bibinfo{person}{Yafeng Chen}, \bibinfo{person}{Yanni Chen}, \bibinfo{person}{Mengzhe Chen}, \bibinfo{person}{Yingda Chen}, \bibinfo{person}{Chong Deng}, \bibinfo{person}{Zhihao Du}, \bibinfo{person}{Ruize Gao}, \bibinfo{person}{Changfeng Gao}, \bibinfo{person}{Zhifu Gao}, {et~al\mbox{.}}} \bibinfo{year}{2025}\natexlab{a}.
\newblock \showarticletitle{Minmo: A multimodal large language model for seamless voice interaction}.
\newblock \bibinfo{journal}{\emph{arXiv preprint arXiv:2501.06282}} (\bibinfo{year}{2025}).
\newblock


\bibitem[Chen et~al\mbox{.}(2024a)]%
        {valle2}
\bibfield{author}{\bibinfo{person}{Sanyuan Chen}, \bibinfo{person}{Shujie Liu}, \bibinfo{person}{Long Zhou}, \bibinfo{person}{Yanqing Liu}, \bibinfo{person}{Xu Tan}, \bibinfo{person}{Jinyu Li}, \bibinfo{person}{Sheng Zhao}, \bibinfo{person}{Yao Qian}, {and} \bibinfo{person}{Furu Wei}.} \bibinfo{year}{2024}\natexlab{a}.
\newblock \showarticletitle{Vall-e 2: Neural codec language models are human parity zero-shot text to speech synthesizers}.
\newblock \bibinfo{journal}{\emph{arXiv preprint arXiv:2406.05370}} (\bibinfo{year}{2024}).
\newblock


\bibitem[Chen et~al\mbox{.}(2025b)]%
        {valle}
\bibfield{author}{\bibinfo{person}{Sanyuan Chen}, \bibinfo{person}{Chengyi Wang}, \bibinfo{person}{Yu Wu}, \bibinfo{person}{Ziqiang Zhang}, \bibinfo{person}{Long Zhou}, \bibinfo{person}{Shujie Liu}, \bibinfo{person}{Zhuo Chen}, \bibinfo{person}{Yanqing Liu}, \bibinfo{person}{Huaming Wang}, \bibinfo{person}{Jinyu Li}, \bibinfo{person}{Lei He}, \bibinfo{person}{Sheng Zhao}, {and} \bibinfo{person}{Furu Wei}.} \bibinfo{year}{2025}\natexlab{b}.
\newblock \showarticletitle{Neural Codec Language Models are Zero-Shot Text to Speech Synthesizers}.
\newblock \bibinfo{journal}{\emph{IEEE Transactions on Audio, Speech and Language Processing}}  \bibinfo{volume}{33} (\bibinfo{year}{2025}), \bibinfo{pages}{705--718}.
\newblock
\href{https://doi.org/10.1109/TASLPRO.2025.3530270}{doi:\nolinkurl{10.1109/TASLPRO.2025.3530270}}


\bibitem[Chen et~al\mbox{.}(2024b)]%
        {3dspeaker}
\bibfield{author}{\bibinfo{person}{Yafeng Chen}, \bibinfo{person}{Siqi Zheng}, \bibinfo{person}{Hui Wang}, \bibinfo{person}{Luyao Cheng}, {et~al\mbox{.}}} \bibinfo{year}{2024}\natexlab{b}.
\newblock \showarticletitle{3D-Speaker-Toolkit: An Open Source Toolkit for Multi-modal Speaker Verification and Diarization}.
\newblock  (\bibinfo{year}{2024}).
\newblock
\urldef\tempurl%
\url{https://arxiv.org/pdf/2403.19971}
\showURL{%
\tempurl}


\bibitem[Cho et~al\mbox{.}(2024)]%
        {cho2024sylber}
\bibfield{author}{\bibinfo{person}{Cheol~Jun Cho}, \bibinfo{person}{Nicholas Lee}, \bibinfo{person}{Akshat Gupta}, \bibinfo{person}{Dhruv Agarwal}, \bibinfo{person}{Ethan Chen}, \bibinfo{person}{Alan~W Black}, {and} \bibinfo{person}{Gopala~K Anumanchipalli}.} \bibinfo{year}{2024}\natexlab{}.
\newblock \showarticletitle{Sylber: Syllabic Embedding Representation of Speech from Raw Audio}.
\newblock \bibinfo{journal}{\emph{arXiv preprint arXiv:2410.07168}} (\bibinfo{year}{2024}).
\newblock


\bibitem[Chu and Peng(2006)]%
        {chu2006objective}
\bibfield{author}{\bibinfo{person}{Min Chu} {and} \bibinfo{person}{Hu Peng}.} \bibinfo{year}{2006}\natexlab{}.
\newblock \bibinfo{title}{Objective measure for estimating mean opinion score of synthesized speech}.
\newblock
\newblock
\shownote{US Patent 7,024,362}.


\bibitem[D{\'e}fossez et~al\mbox{.}(2022)]%
        {encodec}
\bibfield{author}{\bibinfo{person}{Alexandre D{\'e}fossez}, \bibinfo{person}{Jade Copet}, \bibinfo{person}{Gabriel Synnaeve}, {and} \bibinfo{person}{Yossi Adi}.} \bibinfo{year}{2022}\natexlab{}.
\newblock \showarticletitle{High fidelity neural audio compression}.
\newblock \bibinfo{journal}{\emph{arXiv preprint arXiv:2210.13438}} (\bibinfo{year}{2022}).
\newblock


\bibitem[D{\'e}fossez et~al\mbox{.}(2024)]%
        {moshi}
\bibfield{author}{\bibinfo{person}{Alexandre D{\'e}fossez}, \bibinfo{person}{Laurent Mazar{\'e}}, \bibinfo{person}{Manu Orsini}, \bibinfo{person}{Am{\'e}lie Royer}, \bibinfo{person}{Patrick P{\'e}rez}, \bibinfo{person}{Herv{\'e} J{\'e}gou}, \bibinfo{person}{Edouard Grave}, {and} \bibinfo{person}{Neil Zeghidour}.} \bibinfo{year}{2024}\natexlab{}.
\newblock \showarticletitle{Moshi: a speech-text foundation model for real-time dialogue}.
\newblock \bibinfo{journal}{\emph{arXiv preprint arXiv:2410.00037}} (\bibinfo{year}{2024}).
\newblock


\bibitem[Diederik(2014)]%
        {2014adam}
\bibfield{author}{\bibinfo{person}{P~Kingma Diederik}.} \bibinfo{year}{2014}\natexlab{}.
\newblock \showarticletitle{Adam: A method for stochastic optimization}.
\newblock \bibinfo{journal}{\emph{(No Title)}} (\bibinfo{year}{2014}).
\newblock


\bibitem[Dieleman et~al\mbox{.}(2021)]%
        {dieleman2021variable}
\bibfield{author}{\bibinfo{person}{Sander Dieleman}, \bibinfo{person}{Charlie Nash}, \bibinfo{person}{Jesse Engel}, {and} \bibinfo{person}{Karen Simonyan}.} \bibinfo{year}{2021}\natexlab{}.
\newblock \showarticletitle{Variable-rate discrete representation learning}.
\newblock \bibinfo{journal}{\emph{arXiv preprint arXiv:2103.06089}} (\bibinfo{year}{2021}).
\newblock


\bibitem[Du et~al\mbox{.}(2024a)]%
        {cosyvoice}
\bibfield{author}{\bibinfo{person}{Zhihao Du}, \bibinfo{person}{Qian Chen}, \bibinfo{person}{Shiliang Zhang}, \bibinfo{person}{Kai Hu}, \bibinfo{person}{Heng Lu}, \bibinfo{person}{Yexin Yang}, \bibinfo{person}{Hangrui Hu}, \bibinfo{person}{Siqi Zheng}, \bibinfo{person}{Yue Gu}, \bibinfo{person}{Ziyang Ma}, {et~al\mbox{.}}} \bibinfo{year}{2024}\natexlab{a}.
\newblock \showarticletitle{Cosyvoice: A scalable multilingual zero-shot text-to-speech synthesizer based on supervised semantic tokens}.
\newblock \bibinfo{journal}{\emph{arXiv preprint arXiv:2407.05407}} (\bibinfo{year}{2024}).
\newblock


\bibitem[Du et~al\mbox{.}(2024b)]%
        {funcodec}
\bibfield{author}{\bibinfo{person}{Zhihao Du}, \bibinfo{person}{Shiliang Zhang}, \bibinfo{person}{Kai Hu}, {and} \bibinfo{person}{Siqi Zheng}.} \bibinfo{year}{2024}\natexlab{b}.
\newblock \showarticletitle{Funcodec: A fundamental, reproducible and integrable open-source toolkit for neural speech codec}. In \bibinfo{booktitle}{\emph{ICASSP 2024-2024 IEEE International Conference on Acoustics, Speech and Signal Processing (ICASSP)}}. IEEE, \bibinfo{pages}{591--595}.
\newblock


\bibitem[Fu et~al\mbox{.}(2025)]%
        {vita}
\bibfield{author}{\bibinfo{person}{Chaoyou Fu}, \bibinfo{person}{Haojia Lin}, \bibinfo{person}{Xiong Wang}, \bibinfo{person}{Yi-Fan Zhang}, \bibinfo{person}{Yunhang Shen}, \bibinfo{person}{Xiaoyu Liu}, \bibinfo{person}{Yangze Li}, \bibinfo{person}{Zuwei Long}, \bibinfo{person}{Heting Gao}, \bibinfo{person}{Ke Li}, {et~al\mbox{.}}} \bibinfo{year}{2025}\natexlab{}.
\newblock \showarticletitle{Vita-1.5: Towards gpt-4o level real-time vision and speech interaction}.
\newblock \bibinfo{journal}{\emph{arXiv preprint arXiv:2501.01957}} (\bibinfo{year}{2025}).
\newblock


\bibitem[Guo et~al\mbox{.}(2025)]%
        {guo2025recent}
\bibfield{author}{\bibinfo{person}{Yiwei Guo}, \bibinfo{person}{Zhihan Li}, \bibinfo{person}{Hankun Wang}, \bibinfo{person}{Bohan Li}, \bibinfo{person}{Chongtian Shao}, \bibinfo{person}{Hanglei Zhang}, \bibinfo{person}{Chenpeng Du}, \bibinfo{person}{Xie Chen}, \bibinfo{person}{Shujie Liu}, {and} \bibinfo{person}{Kai Yu}.} \bibinfo{year}{2025}\natexlab{}.
\newblock \showarticletitle{Recent Advances in Discrete Speech Tokens: A Review}.
\newblock \bibinfo{journal}{\emph{arXiv preprint arXiv:2502.06490}} (\bibinfo{year}{2025}).
\newblock


\bibitem[Han et~al\mbox{.}(2024)]%
        {valler}
\bibfield{author}{\bibinfo{person}{Bing Han}, \bibinfo{person}{Long Zhou}, \bibinfo{person}{Shujie Liu}, \bibinfo{person}{Sanyuan Chen}, \bibinfo{person}{Lingwei Meng}, \bibinfo{person}{Yanming Qian}, \bibinfo{person}{Yanqing Liu}, \bibinfo{person}{Sheng Zhao}, \bibinfo{person}{Jinyu Li}, {and} \bibinfo{person}{Furu Wei}.} \bibinfo{year}{2024}\natexlab{}.
\newblock \showarticletitle{VALL-E R: Robust and efficient zero-shot text-to-speech synthesis via monotonic alignment}.
\newblock \bibinfo{journal}{\emph{arXiv preprint arXiv:2406.07855}} (\bibinfo{year}{2024}).
\newblock


\bibitem[Hsu et~al\mbox{.}(2021)]%
        {hubert}
\bibfield{author}{\bibinfo{person}{Wei-Ning Hsu}, \bibinfo{person}{Benjamin Bolte}, \bibinfo{person}{Yao-Hung~Hubert Tsai}, \bibinfo{person}{Kushal Lakhotia}, \bibinfo{person}{Ruslan Salakhutdinov}, {and} \bibinfo{person}{Abdelrahman Mohamed}.} \bibinfo{year}{2021}\natexlab{}.
\newblock \showarticletitle{Hubert: Self-supervised speech representation learning by masked prediction of hidden units}.
\newblock \bibinfo{journal}{\emph{IEEE/ACM transactions on audio, speech, and language processing}}  \bibinfo{volume}{29} (\bibinfo{year}{2021}), \bibinfo{pages}{3451--3460}.
\newblock


\bibitem[Ji et~al\mbox{.}(2024)]%
        {wavtokenizer}
\bibfield{author}{\bibinfo{person}{Shengpeng Ji}, \bibinfo{person}{Ziyue Jiang}, \bibinfo{person}{Wen Wang}, \bibinfo{person}{Yifu Chen}, \bibinfo{person}{Minghui Fang}, \bibinfo{person}{Jialong Zuo}, \bibinfo{person}{Qian Yang}, \bibinfo{person}{Xize Cheng}, \bibinfo{person}{Zehan Wang}, \bibinfo{person}{Ruiqi Li}, {et~al\mbox{.}}} \bibinfo{year}{2024}\natexlab{}.
\newblock \showarticletitle{Wavtokenizer: an efficient acoustic discrete codec tokenizer for audio language modeling}.
\newblock \bibinfo{journal}{\emph{arXiv preprint arXiv:2408.16532}} (\bibinfo{year}{2024}).
\newblock


\bibitem[Jiang et~al\mbox{.}(2025)]%
        {unicodec}
\bibfield{author}{\bibinfo{person}{Yidi Jiang}, \bibinfo{person}{Qian Chen}, \bibinfo{person}{Shengpeng Ji}, \bibinfo{person}{Yu Xi}, \bibinfo{person}{Wen Wang}, \bibinfo{person}{Chong Zhang}, \bibinfo{person}{Xianghu Yue}, \bibinfo{person}{ShiLiang Zhang}, {and} \bibinfo{person}{Haizhou Li}.} \bibinfo{year}{2025}\natexlab{}.
\newblock \showarticletitle{UniCodec: Unified Audio Codec with Single Domain-Adaptive Codebook}.
\newblock \bibinfo{journal}{\emph{arXiv preprint arXiv:2502.20067}} (\bibinfo{year}{2025}).
\newblock


\bibitem[Jurafsky et~al\mbox{.}(2008)]%
        {jurafsky2008probabilistic}
\bibfield{author}{\bibinfo{person}{Daniel Jurafsky}, \bibinfo{person}{Alan Bell}, \bibinfo{person}{Michelle Gregory}, {and} \bibinfo{person}{William~D Raymond}.} \bibinfo{year}{2008}\natexlab{}.
\newblock \showarticletitle{Probabilistic relations between words: Evidence from reduction in lexical production}.
\newblock In \bibinfo{booktitle}{\emph{Frequency and the emergence of linguistic structure}}. \bibinfo{publisher}{John Benjamins Publishing Company}, \bibinfo{pages}{229--254}.
\newblock


\bibitem[Kavaki and Mandel(2025)]%
        {kavaki2025audio}
\bibfield{author}{\bibinfo{person}{Hassan~Salami Kavaki} {and} \bibinfo{person}{Michael~I Mandel}.} \bibinfo{year}{2025}\natexlab{}.
\newblock \showarticletitle{Audio Sparse-Transformer for Speech Classification}. In \bibinfo{booktitle}{\emph{ICASSP 2025-2025 IEEE International Conference on Acoustics, Speech and Signal Processing (ICASSP)}}. IEEE, \bibinfo{pages}{1--5}.
\newblock


\bibitem[Kim et~al\mbox{.}(2020a)]%
        {kim2020t}
\bibfield{author}{\bibinfo{person}{Jaeyoung Kim}, \bibinfo{person}{Mostafa El-Khamy}, {and} \bibinfo{person}{Jungwon Lee}.} \bibinfo{year}{2020}\natexlab{a}.
\newblock \showarticletitle{T-gsa: Transformer with gaussian-weighted self-attention for speech enhancement}. In \bibinfo{booktitle}{\emph{ICASSP 2020-2020 IEEE International Conference on Acoustics, Speech and Signal Processing (ICASSP)}}. IEEE, \bibinfo{pages}{6649--6653}.
\newblock


\bibitem[Kim et~al\mbox{.}(2020b)]%
        {glowtts}
\bibfield{author}{\bibinfo{person}{Jaehyeon Kim}, \bibinfo{person}{Sungwon Kim}, \bibinfo{person}{Jungil Kong}, {and} \bibinfo{person}{Sungroh Yoon}.} \bibinfo{year}{2020}\natexlab{b}.
\newblock \showarticletitle{Glow-tts: A generative flow for text-to-speech via monotonic alignment search}.
\newblock \bibinfo{journal}{\emph{Advances in Neural Information Processing Systems}}  \bibinfo{volume}{33} (\bibinfo{year}{2020}), \bibinfo{pages}{8067--8077}.
\newblock


\bibitem[Kong et~al\mbox{.}(2020)]%
        {hifigan}
\bibfield{author}{\bibinfo{person}{Jungil Kong}, \bibinfo{person}{Jaehyeon Kim}, {and} \bibinfo{person}{Jaekyoung Bae}.} \bibinfo{year}{2020}\natexlab{}.
\newblock \showarticletitle{Hifi-gan: Generative adversarial networks for efficient and high fidelity speech synthesis}.
\newblock \bibinfo{journal}{\emph{Advances in neural information processing systems}}  \bibinfo{volume}{33} (\bibinfo{year}{2020}), \bibinfo{pages}{17022--17033}.
\newblock


\bibitem[Li et~al\mbox{.}(2025)]%
        {baichuanomni}
\bibfield{author}{\bibinfo{person}{Yadong Li}, \bibinfo{person}{Jun Liu}, \bibinfo{person}{Tao Zhang}, \bibinfo{person}{Song Chen}, \bibinfo{person}{Tianpeng Li}, \bibinfo{person}{Zehuan Li}, \bibinfo{person}{Lijun Liu}, \bibinfo{person}{Lingfeng Ming}, \bibinfo{person}{Guosheng Dong}, \bibinfo{person}{Da Pan}, {et~al\mbox{.}}} \bibinfo{year}{2025}\natexlab{}.
\newblock \showarticletitle{Baichuan-Omni-1.5 Technical Report}.
\newblock \bibinfo{journal}{\emph{arXiv preprint arXiv:2501.15368}} (\bibinfo{year}{2025}).
\newblock


\bibitem[Lu et~al\mbox{.}(2025)]%
        {lu2025moba}
\bibfield{author}{\bibinfo{person}{Enzhe Lu}, \bibinfo{person}{Zhejun Jiang}, \bibinfo{person}{Jingyuan Liu}, \bibinfo{person}{Yulun Du}, \bibinfo{person}{Tao Jiang}, \bibinfo{person}{Chao Hong}, \bibinfo{person}{Shaowei Liu}, \bibinfo{person}{Weiran He}, \bibinfo{person}{Enming Yuan}, \bibinfo{person}{Yuzhi Wang}, {et~al\mbox{.}}} \bibinfo{year}{2025}\natexlab{}.
\newblock \showarticletitle{MoBA: Mixture of Block Attention for Long-Context LLMs}.
\newblock \bibinfo{journal}{\emph{arXiv preprint arXiv:2502.13189}} (\bibinfo{year}{2025}).
\newblock


\bibitem[Malisz et~al\mbox{.}(2018)]%
        {malisz2018dimensions}
\bibfield{author}{\bibinfo{person}{Zofia Malisz}, \bibinfo{person}{Erika Brandt}, \bibinfo{person}{Bernd M{\"o}bius}, \bibinfo{person}{Yoon~Mi Oh}, {and} \bibinfo{person}{Bistra Andreeva}.} \bibinfo{year}{2018}\natexlab{}.
\newblock \showarticletitle{Dimensions of segmental variability: Interaction of prosody and surprisal in six languages}.
\newblock \bibinfo{journal}{\emph{Frontiers in Communication}}  \bibinfo{volume}{3} (\bibinfo{year}{2018}), \bibinfo{pages}{25}.
\newblock


\bibitem[Nishimura et~al\mbox{.}(2024)]%
        {halle}
\bibfield{author}{\bibinfo{person}{Yuto Nishimura}, \bibinfo{person}{Takumi Hirose}, \bibinfo{person}{Masanari Ohi}, \bibinfo{person}{Hideki Nakayama}, {and} \bibinfo{person}{Nakamasa Inoue}.} \bibinfo{year}{2024}\natexlab{}.
\newblock \showarticletitle{HALL-E: hierarchical neural codec language model for minute-long zero-shot text-to-speech synthesis}.
\newblock \bibinfo{journal}{\emph{arXiv preprint arXiv:2410.04380}} (\bibinfo{year}{2024}).
\newblock


\bibitem[Panayotov et~al\mbox{.}(2015)]%
        {librispeech}
\bibfield{author}{\bibinfo{person}{Vassil Panayotov}, \bibinfo{person}{Guoguo Chen}, \bibinfo{person}{Daniel Povey}, {and} \bibinfo{person}{Sanjeev Khudanpur}.} \bibinfo{year}{2015}\natexlab{}.
\newblock \showarticletitle{Librispeech: An ASR corpus based on public domain audio books}. In \bibinfo{booktitle}{\emph{2015 IEEE International Conference on Acoustics, Speech and Signal Processing (ICASSP)}}. \bibinfo{pages}{5206--5210}.
\newblock
\href{https://doi.org/10.1109/ICASSP.2015.7178964}{doi:\nolinkurl{10.1109/ICASSP.2015.7178964}}


\bibitem[Peng et~al\mbox{.}(2024)]%
        {voicecraft}
\bibfield{author}{\bibinfo{person}{Puyuan Peng}, \bibinfo{person}{Po-Yao Huang}, \bibinfo{person}{Shang-Wen Li}, \bibinfo{person}{Abdelrahman Mohamed}, {and} \bibinfo{person}{David Harwath}.} \bibinfo{year}{2024}\natexlab{}.
\newblock \showarticletitle{Voicecraft: Zero-shot speech editing and text-to-speech in the wild}.
\newblock \bibinfo{journal}{\emph{arXiv preprint arXiv:2403.16973}} (\bibinfo{year}{2024}).
\newblock


\bibitem[Pratap et~al\mbox{.}(2020)]%
        {MLS}
\bibfield{author}{\bibinfo{person}{Vineel Pratap}, \bibinfo{person}{Qiantong Xu}, \bibinfo{person}{Anuroop Sriram}, \bibinfo{person}{Gabriel Synnaeve}, {and} \bibinfo{person}{Ronan Collobert}.} \bibinfo{year}{2020}\natexlab{}.
\newblock \showarticletitle{Mls: A large-scale multilingual dataset for speech research}.
\newblock \bibinfo{journal}{\emph{arXiv preprint arXiv:2012.03411}} (\bibinfo{year}{2020}).
\newblock


\bibitem[Radford et~al\mbox{.}(2023)]%
        {whisper}
\bibfield{author}{\bibinfo{person}{Alec Radford}, \bibinfo{person}{Jong~Wook Kim}, \bibinfo{person}{Tao Xu}, \bibinfo{person}{Greg Brockman}, \bibinfo{person}{Christine McLeavey}, {and} \bibinfo{person}{Ilya Sutskever}.} \bibinfo{year}{2023}\natexlab{}.
\newblock \showarticletitle{Robust speech recognition via large-scale weak supervision}. In \bibinfo{booktitle}{\emph{International conference on machine learning}}. PMLR, \bibinfo{pages}{28492--28518}.
\newblock


\bibitem[Ren et~al\mbox{.}(2024)]%
        {ticodec}
\bibfield{author}{\bibinfo{person}{Yong Ren}, \bibinfo{person}{Tao Wang}, \bibinfo{person}{Jiangyan Yi}, \bibinfo{person}{Le Xu}, \bibinfo{person}{Jianhua Tao}, \bibinfo{person}{Chu~Yuan Zhang}, {and} \bibinfo{person}{Junzuo Zhou}.} \bibinfo{year}{2024}\natexlab{}.
\newblock \showarticletitle{Fewer-token neural speech codec with time-invariant codes}. In \bibinfo{booktitle}{\emph{ICASSP 2024-2024 IEEE International Conference on Acoustics, Speech and Signal Processing (ICASSP)}}. IEEE, \bibinfo{pages}{12737--12741}.
\newblock


\bibitem[Saeki et~al\mbox{.}(2022)]%
        {utmos}
\bibfield{author}{\bibinfo{person}{Takaaki Saeki}, \bibinfo{person}{Detai Xin}, \bibinfo{person}{Wataru Nakata}, \bibinfo{person}{Tomoki Koriyama}, \bibinfo{person}{Shinnosuke Takamichi}, {and} \bibinfo{person}{Hiroshi Saruwatari}.} \bibinfo{year}{2022}\natexlab{}.
\newblock \showarticletitle{Utmos: Utokyo-sarulab system for voicemos challenge 2022}.
\newblock \bibinfo{journal}{\emph{arXiv preprint arXiv:2204.02152}} (\bibinfo{year}{2022}).
\newblock


\bibitem[Shi et~al\mbox{.}(2023)]%
        {shi2023large}
\bibfield{author}{\bibinfo{person}{Freda Shi}, \bibinfo{person}{Xinyun Chen}, \bibinfo{person}{Kanishka Misra}, \bibinfo{person}{Nathan Scales}, \bibinfo{person}{David Dohan}, \bibinfo{person}{Ed~H Chi}, \bibinfo{person}{Nathanael Sch{\"a}rli}, {and} \bibinfo{person}{Denny Zhou}.} \bibinfo{year}{2023}\natexlab{}.
\newblock \showarticletitle{Large language models can be easily distracted by irrelevant context}. In \bibinfo{booktitle}{\emph{International Conference on Machine Learning}}. PMLR, \bibinfo{pages}{31210--31227}.
\newblock


\bibitem[Shih et~al\mbox{.}(2021)]%
        {radtts}
\bibfield{author}{\bibinfo{person}{Kevin~J Shih}, \bibinfo{person}{Rafael Valle}, \bibinfo{person}{Rohan Badlani}, \bibinfo{person}{Adrian Lancucki}, \bibinfo{person}{Wei Ping}, {and} \bibinfo{person}{Bryan Catanzaro}.} \bibinfo{year}{2021}\natexlab{}.
\newblock \showarticletitle{RAD-TTS: Parallel flow-based TTS with robust alignment learning and diverse synthesis}. In \bibinfo{booktitle}{\emph{ICML Workshop on Invertible Neural Networks, Normalizing Flows, and Explicit Likelihood Models}}.
\newblock


\bibitem[Sicherman and Adi(2023)]%
        {sicherman2023analysing}
\bibfield{author}{\bibinfo{person}{Amitay Sicherman} {and} \bibinfo{person}{Yossi Adi}.} \bibinfo{year}{2023}\natexlab{}.
\newblock \showarticletitle{Analysing discrete self supervised speech representation for spoken language modeling}. In \bibinfo{booktitle}{\emph{ICASSP 2023-2023 IEEE International Conference on Acoustics, Speech and Signal Processing (ICASSP)}}. IEEE, \bibinfo{pages}{1--5}.
\newblock


\bibitem[Song et~al\mbox{.}(2024)]%
        {ellav}
\bibfield{author}{\bibinfo{person}{Yakun Song}, \bibinfo{person}{Zhuo Chen}, \bibinfo{person}{Xiaofei Wang}, \bibinfo{person}{Ziyang Ma}, {and} \bibinfo{person}{Xie Chen}.} \bibinfo{year}{2024}\natexlab{}.
\newblock \showarticletitle{Ella-v: Stable neural codec language modeling with alignment-guided sequence reordering}.
\newblock \bibinfo{journal}{\emph{arXiv preprint arXiv:2401.07333}} (\bibinfo{year}{2024}).
\newblock


\bibitem[Tikochinski et~al\mbox{.}(2025)]%
        {tikochinski2025incremental}
\bibfield{author}{\bibinfo{person}{Refael Tikochinski}, \bibinfo{person}{Ariel Goldstein}, \bibinfo{person}{Yoav Meiri}, \bibinfo{person}{Uri Hasson}, {and} \bibinfo{person}{Roi Reichart}.} \bibinfo{year}{2025}\natexlab{}.
\newblock \showarticletitle{Incremental accumulation of linguistic context in artificial and biological neural networks}.
\newblock \bibinfo{journal}{\emph{Nature Communications}} \bibinfo{volume}{16}, \bibinfo{number}{1} (\bibinfo{year}{2025}), \bibinfo{pages}{803}.
\newblock


\bibitem[Touvron et~al\mbox{.}(2023)]%
        {2023llama}
\bibfield{author}{\bibinfo{person}{Hugo Touvron}, \bibinfo{person}{Thibaut Lavril}, \bibinfo{person}{Gautier Izacard}, \bibinfo{person}{Xavier Martinet}, \bibinfo{person}{Marie-Anne Lachaux}, \bibinfo{person}{Timoth{\'e}e Lacroix}, \bibinfo{person}{Baptiste Rozi{\`e}re}, \bibinfo{person}{Naman Goyal}, \bibinfo{person}{Eric Hambro}, \bibinfo{person}{Faisal Azhar}, {et~al\mbox{.}}} \bibinfo{year}{2023}\natexlab{}.
\newblock \showarticletitle{Llama: Open and efficient foundation language models}.
\newblock \bibinfo{journal}{\emph{arXiv preprint arXiv:2302.13971}} (\bibinfo{year}{2023}).
\newblock


\bibitem[Wang et~al\mbox{.}(2025)]%
        {wang2025vocalnetspeechllmmultitoken}
\bibfield{author}{\bibinfo{person}{Yuhao Wang}, \bibinfo{person}{Heyang Liu}, \bibinfo{person}{Ziyang Cheng}, \bibinfo{person}{Ronghua Wu}, \bibinfo{person}{Qunshan Gu}, \bibinfo{person}{Yanfeng Wang}, {and} \bibinfo{person}{Yu Wang}.} \bibinfo{year}{2025}\natexlab{}.
\newblock \bibinfo{title}{VocalNet: Speech LLM with Multi-Token Prediction for Faster and High-Quality Generation}.
\newblock
\showeprint[arxiv]{2504.04060}~[cs.CL]
\urldef\tempurl%
\url{https://arxiv.org/abs/2504.04060}
\showURL{%
\tempurl}


\bibitem[Wang et~al\mbox{.}(2024)]%
        {maskgct}
\bibfield{author}{\bibinfo{person}{Yuancheng Wang}, \bibinfo{person}{Haoyue Zhan}, \bibinfo{person}{Liwei Liu}, \bibinfo{person}{Ruihong Zeng}, \bibinfo{person}{Haotian Guo}, \bibinfo{person}{Jiachen Zheng}, \bibinfo{person}{Qiang Zhang}, \bibinfo{person}{Xueyao Zhang}, \bibinfo{person}{Shunsi Zhang}, {and} \bibinfo{person}{Zhizheng Wu}.} \bibinfo{year}{2024}\natexlab{}.
\newblock \showarticletitle{Maskgct: Zero-shot text-to-speech with masked generative codec transformer}.
\newblock \bibinfo{journal}{\emph{arXiv preprint arXiv:2409.00750}} (\bibinfo{year}{2024}).
\newblock


\bibitem[Xin et~al\mbox{.}(2024)]%
        {bigcodec}
\bibfield{author}{\bibinfo{person}{Detai Xin}, \bibinfo{person}{Xu Tan}, \bibinfo{person}{Shinnosuke Takamichi}, {and} \bibinfo{person}{Hiroshi Saruwatari}.} \bibinfo{year}{2024}\natexlab{}.
\newblock \showarticletitle{Bigcodec: Pushing the limits of low-bitrate neural speech codec}.
\newblock \bibinfo{journal}{\emph{arXiv preprint arXiv:2409.05377}} (\bibinfo{year}{2024}).
\newblock


\bibitem[Yang et~al\mbox{.}(2023)]%
        {hificodec}
\bibfield{author}{\bibinfo{person}{Dongchao Yang}, \bibinfo{person}{Songxiang Liu}, \bibinfo{person}{Rongjie Huang}, \bibinfo{person}{Jinchuan Tian}, \bibinfo{person}{Chao Weng}, {and} \bibinfo{person}{Yuexian Zou}.} \bibinfo{year}{2023}\natexlab{}.
\newblock \showarticletitle{Hifi-codec: Group-residual vector quantization for high fidelity audio codec}.
\newblock \bibinfo{journal}{\emph{arXiv preprint arXiv:2305.02765}} (\bibinfo{year}{2023}).
\newblock


\bibitem[Yang et~al\mbox{.}(2024)]%
        {uniaudio}
\bibfield{author}{\bibinfo{person}{Dongchao Yang}, \bibinfo{person}{Jinchuan Tian}, \bibinfo{person}{Xu Tan}, \bibinfo{person}{Rongjie Huang}, \bibinfo{person}{Songxiang Liu}, \bibinfo{person}{Haohan Guo}, \bibinfo{person}{Xuankai Chang}, \bibinfo{person}{Jiatong Shi}, \bibinfo{person}{Jiang Bian}, \bibinfo{person}{Zhou Zhao}, {et~al\mbox{.}}} \bibinfo{year}{2024}\natexlab{}.
\newblock \showarticletitle{Uniaudio: Towards universal audio generation with large language models}. In \bibinfo{booktitle}{\emph{Forty-first International Conference on Machine Learning}}.
\newblock


\bibitem[Ye et~al\mbox{.}(2024)]%
        {xcodec}
\bibfield{author}{\bibinfo{person}{Zhen Ye}, \bibinfo{person}{Peiwen Sun}, \bibinfo{person}{Jiahe Lei}, \bibinfo{person}{Hongzhan Lin}, \bibinfo{person}{Xu Tan}, \bibinfo{person}{Zheqi Dai}, \bibinfo{person}{Qiuqiang Kong}, \bibinfo{person}{Jianyi Chen}, \bibinfo{person}{Jiahao Pan}, \bibinfo{person}{Qifeng Liu}, {et~al\mbox{.}}} \bibinfo{year}{2024}\natexlab{}.
\newblock \showarticletitle{Codec does matter: Exploring the semantic shortcoming of codec for audio language model}.
\newblock \bibinfo{journal}{\emph{arXiv preprint arXiv:2408.17175}} (\bibinfo{year}{2024}).
\newblock


\bibitem[Yu et~al\mbox{.}(2023)]%
        {megabyte}
\bibfield{author}{\bibinfo{person}{Lili Yu}, \bibinfo{person}{D{\'a}niel Simig}, \bibinfo{person}{Colin Flaherty}, \bibinfo{person}{Armen Aghajanyan}, \bibinfo{person}{Luke Zettlemoyer}, {and} \bibinfo{person}{Mike Lewis}.} \bibinfo{year}{2023}\natexlab{}.
\newblock \showarticletitle{Megabyte: Predicting million-byte sequences with multiscale transformers}.
\newblock \bibinfo{journal}{\emph{Advances in Neural Information Processing Systems}}  \bibinfo{volume}{36} (\bibinfo{year}{2023}), \bibinfo{pages}{78808--78823}.
\newblock


\bibitem[Zeghidour et~al\mbox{.}(2021)]%
        {soundstream}
\bibfield{author}{\bibinfo{person}{Neil Zeghidour}, \bibinfo{person}{Alejandro Luebs}, \bibinfo{person}{Ahmed Omran}, \bibinfo{person}{Jan Skoglund}, {and} \bibinfo{person}{Marco Tagliasacchi}.} \bibinfo{year}{2021}\natexlab{}.
\newblock \showarticletitle{Soundstream: An end-to-end neural audio codec}.
\newblock \bibinfo{journal}{\emph{IEEE/ACM Transactions on Audio, Speech, and Language Processing}}  \bibinfo{volume}{30} (\bibinfo{year}{2021}), \bibinfo{pages}{495--507}.
\newblock


\bibitem[Zen et~al\mbox{.}(2019)]%
        {libritts}
\bibfield{author}{\bibinfo{person}{Heiga Zen}, \bibinfo{person}{Viet Dang}, \bibinfo{person}{Rob Clark}, \bibinfo{person}{Yu Zhang}, \bibinfo{person}{Ron~J. Weiss}, \bibinfo{person}{Ye Jia}, \bibinfo{person}{Zhifeng Chen}, {and} \bibinfo{person}{Yonghui Wu}.} \bibinfo{year}{2019}\natexlab{}.
\newblock \bibinfo{title}{LibriTTS: A Corpus Derived from LibriSpeech for Text-to-Speech}.
\newblock
\showeprint[arxiv]{1904.02882}~[cs.SD]
\urldef\tempurl%
\url{https://arxiv.org/abs/1904.02882}
\showURL{%
\tempurl}


\bibitem[Zhang et~al\mbox{.}(2023a)]%
        {speechgpt}
\bibfield{author}{\bibinfo{person}{Dong Zhang}, \bibinfo{person}{Shimin Li}, \bibinfo{person}{Xin Zhang}, \bibinfo{person}{Jun Zhan}, \bibinfo{person}{Pengyu Wang}, \bibinfo{person}{Yaqian Zhou}, {and} \bibinfo{person}{Xipeng Qiu}.} \bibinfo{year}{2023}\natexlab{a}.
\newblock \showarticletitle{Speechgpt: Empowering large language models with intrinsic cross-modal conversational abilities}.
\newblock \bibinfo{journal}{\emph{arXiv preprint arXiv:2305.11000}} (\bibinfo{year}{2023}).
\newblock


\bibitem[Zhang et~al\mbox{.}(2023b)]%
        {speechtokenizer}
\bibfield{author}{\bibinfo{person}{Xin Zhang}, \bibinfo{person}{Dong Zhang}, \bibinfo{person}{Shimin Li}, \bibinfo{person}{Yaqian Zhou}, {and} \bibinfo{person}{Xipeng Qiu}.} \bibinfo{year}{2023}\natexlab{b}.
\newblock \showarticletitle{Speechtokenizer: Unified speech tokenizer for speech large language models}.
\newblock \bibinfo{journal}{\emph{arXiv preprint arXiv:2308.16692}} (\bibinfo{year}{2023}).
\newblock


\end{thebibliography}

\clearpage
\appendix
\section{Appendix}
\subsection{More Details about Method}

\begin{figure*}[t!]
    \centering
    \includegraphics[scale=0.38]{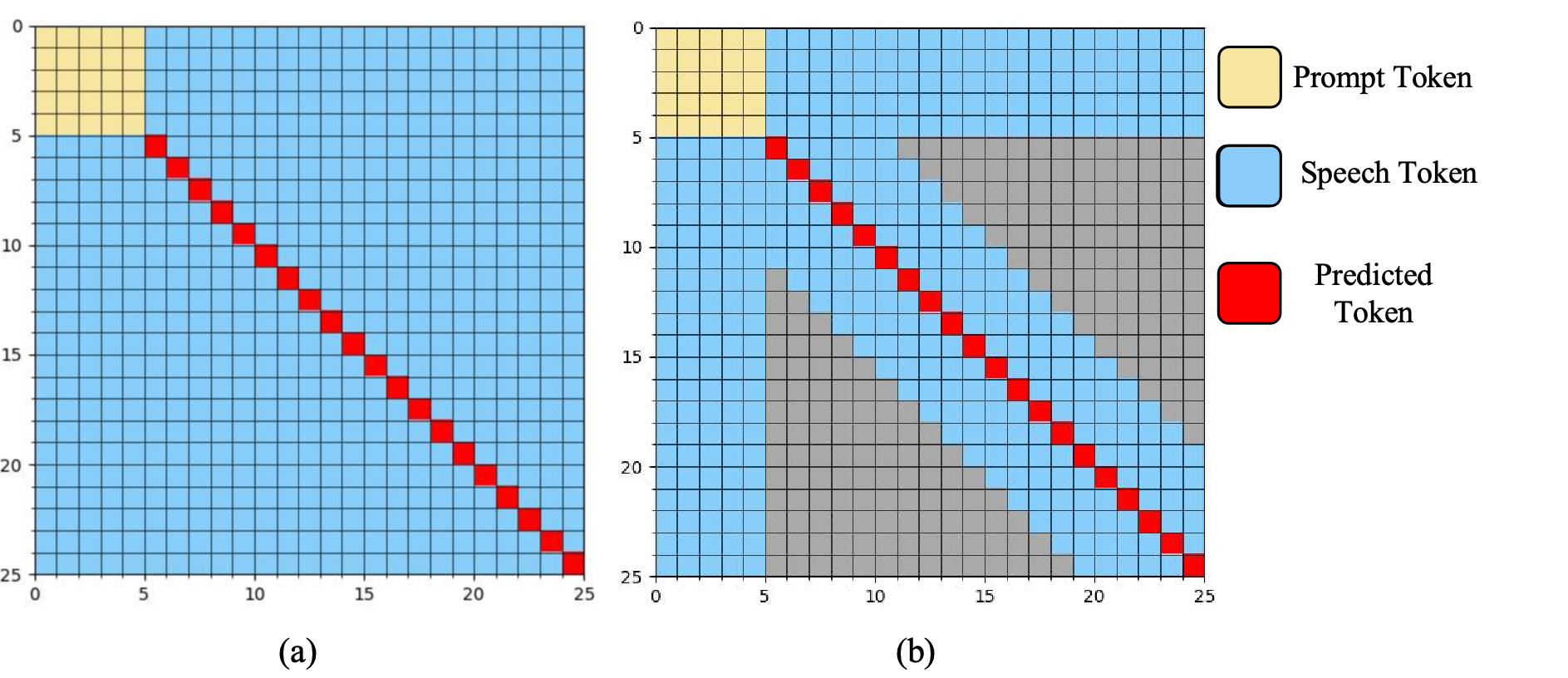}
    \caption{Illustration of the Dense Attention and Prompt-Local Sliding Window Attention in Non-Autogressive Language Models. (a) In \textit{dense attention}, the current speech token can attend to all codewords from previous quantizers. (b) In \textit{prompt-local sliding window attention}, the current speech token can only attend to prompt tokens and codewords from previous quantizers within the bidirectional local sliding window.}
    \label{fig:nar_attention}
\end{figure*}

\subsection{Pseudo Code about Faster Inference}
\begin{algorithm}
  \SetAlgoLined
  \KwData{
    prompt tokens: $[T, C^{\text{ref}}]$, \\
    the number of decoder layers in $\text{LM}_{\text{AR}}$: $J$,\\
    output head in the last decoder layer: $\text{LM}_{\text{AR}}^{\text{head}}$, \\
    local window size: $N$, \\
    Compressed token: $W$, \\
    Interval for inserting $W$: $G$, \\
    begin token: $\text{BOS}$, \\
    end token: $\text{EOS}$, \\
    }
  \KwResult{predicted speech tokens $C^{\text{pred}}$, KV-Cache $K$ and $V$}

  initialization\;
    input sequence: $X \leftarrow [T, C^{\text{ref}}, \text{BOS}]$ \;
    number of prompt tokens: $N_{\text{p}} = \text{len}(X)$ \;
    KV cache: $K \leftarrow \phi$, $V \leftarrow \phi$ \;
    KV cache for compressed tokens: $K_W \leftarrow \phi$, $V_W \leftarrow \phi$ \;
    predicted speech tokens: $C^{pred} \leftarrow \phi$ \;
    $t \leftarrow 0$ \;
  \While{True}
  {
    \eIf{$t == 0$ or $t \% G \neq 0$}
    {
        Modify the attention mask so that $C^{\text{pred}}_{0:-(N-1)}$ is not visible. \;
        $h, K_t, V_t = \text{LM}_{\text{AR}}^{1:J}(X, K, V, t)$ \;
        \If{$t > N$}
        {
            $K \leftarrow [K_{:N_{\text{p}}}, K_W, K_{-(N - 1):}]$ \;
            $V \leftarrow [V_{:N_{\text{p}}}, V_W, V_{-(N - 1):}]$ \;
        }
        $K \leftarrow \text{concat}(K, K_t)$ \;
        $V \leftarrow \text{concat}(V, V_t)$ \;
        $C_t = \text{sample}(\text{softmax}(\text{LM}_{\text{AR}}^{\text{head}}(h)))$ \;
        $C^{pred} \leftarrow {C^{pred} \cup \{C_t\}}$ \;
        \If{$C_t == \text{EOS}$}
        {
            break \;
        }
    }
    {
        Modify the attention mask so that $W$ can only attend to $C^{\text{pred}}_{-G:}$. \;
        $k = \frac{t}{G}$ \;
        $W_k, K_t, V_t = \text{LM}_{\text{AR}}^{1:J}(W, K, V, t)$ \;
        $K_W \leftarrow K_W \cup K_t$ \;
        $V_W \leftarrow V_W \cup V_t$ \;
    }
    $t \leftarrow t + 1$ \;
    }
  \caption{Accelerating $\text{LM}_{\text{AR}}$ Inference with Compressed-to-fine Language Modeling}
  \label{inference}
\end{algorithm}


\end{document}